\documentclass[12pt]{article}

\usepackage{amsmath}
\usepackage{graphicx}
\usepackage{enumerate}
\usepackage{natbib}
\usepackage{url}
\usepackage{color}

\usepackage{bm}
\bmdefine{\btheta}{\theta}

\usepackage[ruled,vlined]{algorithm2e}
\SetKwInput{KwInput}{Input}
\SetKwInput{KwOutput}{Output}

\begin{document}

\def\spacingset#1{\renewcommand{\baselinestretch}%
{#1}\small\normalsize} \spacingset{1}

\thispagestyle{plain}
\begin{center}
\Large{\textbf{Integrating Region-Specific SARS-CoV-2 Data for Statistical Wastewater Monitoring}}
       
\vspace{0.4cm}
Anastasios Apsemidis$^{(a)}$, Karin Weyermair$^{(b)}$, Hans Peter Stüger$^{(b)}$, Sabrina Kuchling$^{(b)}$, Tadej Zerak$^{(b)}$ and Oliver Alber$^{(b)}$
\vspace{0.4cm}

\normalsize{$^{(a)}$Department of Primary Education, University of Ioannina, Ioannina, Greece, {\tt a.apsemidis@uoi.gr}} \\
\normalsize{$^{(b)}$Department Statistics and Analytical Epidemiology, Austrian Agency for Health and Food Safety, Zinzendorfgasse 27, 8010 Graz, Austria, {\tt karin.weyermair@ages.at}, {\tt hans-peter.stueger@ages.at}, {\tt sabrina.kuchling@ages.at}, {\tt tadej.zerak@ages.at}, {\tt oliver.alber@ages.at}} \\
\end{center}

\bigskip
\begin{abstract}
Wastewater data can be very useful for epidemic control during a disease outbreak and proper synthesis of different sources of information can be integrated towards an alerting system, that can be used for decision support. Wastewater data are considered to be of high quality, since they do not depend on testing and can take into account asymptomatic cases. However, little effort has been given into utilizing such information in statistical process control procedures, usually aimed at industrial problems. In this article, we demonstrate how wastewater data can be utilized in health surveillance and propose a statistical framework that can act as a decision support tool. Specifically, we analyze SARS-CoV-2 wastewater data from Austria, constructing summary variables to implicitly describe the Covid-19 prevalence and, based on them, we assess the effectiveness of the current sampling strategy of Austria. We propose a framework of a statistical process monitoring system to aid epidemic management procedures in case SARS-CoV-2 concentration gets dangerously high.
\end{abstract}

\noindent%
\vfill

\newpage
\spacingset{1.9}
\section{Introduction}
\label{sec:intro}

Wastewater monitoring is a relatively novel method of estimating the concentration of viruses in a community, which became a routine after the Severe Acute Respiratory Syndrome Coronavirus 2 (SARS-CoV-2) outbreak in 2019. The complexity of the epidemiological characteristics of the disease indicate that direct measurements such as the number of daily cases do not suffice to estimate the prevalence of the Coronavirus disease 2019 (Covid-19). On the other hand, analysis of SARS-CoV-2 concentration in the wastewater provides a more accurate method, since such samples contain information from the whole population and not only the ones tested (see for example \citealt{cevik2021sars}), giving rise to the wastewater-based epidemiology (WBE). The information flow in epidemic data is usually represented by a pyramid (see for instance \citealt{medema2020implementation}, or \citealt{gibbons2014measuring}), whose base is occupied by the total cases (symptomatic and asymptomatic) and, as we move to higher (and narrower) sectors, fewer individuals are observed (for example deaths occupy the top sector of the pyramid). In this context, wastewater measurements provide access to lower parts of the pyramid maintaining high quality information.

Current uses of wastewater data range from studies from an epidemiological point of view, to statistical analyses and modeling (e.g. \citealt{somerset2024wastewater}, or \citealt{pajaro2022stochastic}). For example, \citet{teunis2015shedding} study the time course of shedding of norovirus to the wastewater, while \citet{miura2021duration} use the same model to estimate the duration of SARS-CoV-2 shedding to be approximately 26 days from symptom onset and its concentration in faeces to be 2.6 log-copies/g. \citet{riedmann2025estimates} use wastewater data to estimate the number of SARS-CoV-2 infections in Austria, as well as the immunity levels of the population (see also \citealt{rauch2024estimating}), while analyses on Greek data are available from \citet{kostoglou2024effect} and \citet{koureas2021wastewater}. A review for SARS-CoV-2 WBE can be found in \citet{zhang2022sars} and in \citet{ciannella2023recent}.

In the present article, we focus on SARS-CoV-2 wastewater measurements obtained from 48 municipal wastewater treatment plants (WWTPs) throughout Austria. We aim at summarizing this region-specific information to the national level, assessing the uncertainty involved and constructing an automated alerting system for epidemic management. Our final goal is to propose some ideas that can be used by a country that gathers wastewater information and thus enhance the current decision making procedures during an epidemic outbreak. To this end, we propose two aggregated national statistics and show three examples of their usage. The first is a retrospective assessment of `sub-sampling' scenarios, quantifying the similarity of the original national wastewater sampling program to a less costly one. Then, we propose the introduction of statistical process control and monitoring ideas in the WBE context and, finally we show that training a model on these statistics is possible and has a natural interpretation.

The remaining of the article is organized as follows. In Section \ref{sec:motiv} we present the data and the motivating real example for this research, as well as the construction of two wastewater national curves. Section \ref{sec:samstraeval} describes the usage of the constructed statistics in evaluating current sampling strategies, while in Section \ref{sec:spm} we use them for an automated alerting system. Section \ref{sec:modbuild} describes the use of the national curves in model building providing an illustration. Section \ref{sec:discuss} concludes with a discussion and ideas for future research.

\section{Data and motivating application}
\label{sec:motiv}

We have measurements from 48 WWTPs in Austria during 2023-2024, each containing twice-per-week samples of the virus concentration (measured in RNA copies per ml) and the inflow of the wastewater in the sewer (measured in cubic meters per day). The full dataset (see Appendix A) contains much more information, but these two provide the foundation of the monitoring system.

It has been shown that many factors can influence the virus concentration measurements (see for example \citealt{bertels2022factors}, or \citealt{rose2015characterization} for a review), so analysis of such data can be considered in the future, but we do not pursue such kind of analysis in the present article. Therefore, instead of the concentration, we propose using the variable
\begin{equation} \label{eq:monit}
E_{t,w} = \frac{c_{t,w} I_{t,w} 10^6}{16\cdot 10^9}
\end{equation}
which represents the number of `fictitious' excretors. This is an estimate of the infected population size at time $t$ in the population of a specific catchment area $w$. We have denoted as $c_{t,w}$ the virus concentration and as $I_{t,w}$ the inflow in the sewer. The factor $10^6$ converts the $\mathrm{ml}$ of $c_{t,w}$ to $\mathrm{L}$ and the $\mathrm{m^3}$ of $I_{t,w}$ to $\mathrm{L}$. Then, the variable $V_t=c_{t,w} I_{t,w} 10^6$ is called the virus load and it is expressed as RNA copies per day. The denominator of $E_{t,w}$ is the estimated average viral excretion per infected person and day.

Since we deal with panel data, the conversion to time series involves some kind of imputation and, so we apply a linear interpolation and refer to the imputed variable as the fictitious excretors time series from 19/01/2023 until 31/12/2024.

\subsection{Aggregation methods}

The monitoring variable in equation \ref{eq:monit} refers to daily measurements at each of 48 treatment plants, which we wish to aggregate to one dynamic statistic for Austria. To this end, we use two different methods. First, we construct the number of the fictitious excretors per 100,000 residents as
\begin{equation*}
Y_t=10^5\frac{E_{t,w}}{R_{t,w}}
\end{equation*}
where $R_{t,w}$ is the population size at the catchment area of WWTP $w$ at day $t$. Then, the two aggregation methods use
\begin{align*}
Y_t^{(1)} &= 10^5\frac{\sum_wE_{t,w}}{\sum_wR_{t,w}} \\
Y_t^{(2)} &= Q_t\Big(10^5\frac{E_{t,w}}{R_{t,w}}\Big)
\end{align*}
where $Q(\cdot)$ is the quantile function (we use as default the median) over samples at each day. The two methods give very similar results (also see Appendix B). The method-1 statistic aims at directly summarizing the whole dataset, while the method-2 statistic uses a `sampling approach', since it deals with measurements of each day as draws from the national-level curve. The inherent variability of each statistic, as well as the influence that each WWTP has on the final national curve are studies in Appendices C and D respectively).

\section{Sampling strategies evaluation}
\label{sec:samstraeval}

In this Section, we investigate whether we can switch from the current (costly) sampling scheme of sampling twice per week at 48 WWTP's to a less frequent one and still obtain accurate results. Thus, the current sampling system serves as a reference, which produces a certain national curve and we would like to reduce either sampling volume or frequency without great loss of information.

To this end, we construct 11 `theoretical' sampling scenarios, one from each combination of volume and frequency (also see Appendix E). Regarding sampling volume, the options are using only 1 WWTP (the largest with respect to the population size in the catchment area), 9 WWTP's (the largest in each state), 24 WWTP's (the ones initially in the program), or 48 WWTP's (excluding none). Regarding sampling frequency, the options are sampling only once per two weeks (keeping the first measurement during two weeks), once per week (excluding one sample) or twice per week (excluding none).

For each combination, we constructed the aggregated monitoring variable as if we only had this information available and then compared with the original national curve using some measure of dissimilarity. We use the $L_2$ distance also known as Euclidean norm, correlation and cross-correlation -based dissimilarities for time series, calculated as $L_2=\sqrt{\sum_{i=1}^N(x_i-y_i)^2}$, $C_1=2(1-r)$ and $C_2=\sqrt{(1-r)/(\sum_{k=1}^K CC_k)}$ respectively. We have denoted as $x_i$ and $y_i$ the two time series at time $i=1,...,N$, $r$ is the correlation and $CC_k$ is the cross-correlation at lag $k=1,...,K$. The results using method-1 and the correlation-based dissimilarity are shown in Figure \ref{fig:corsimmth1} (the rest can be found in Appendix E).

\begin{figure}[h!]
\begin{center}
\includegraphics[width=0.9\textwidth]{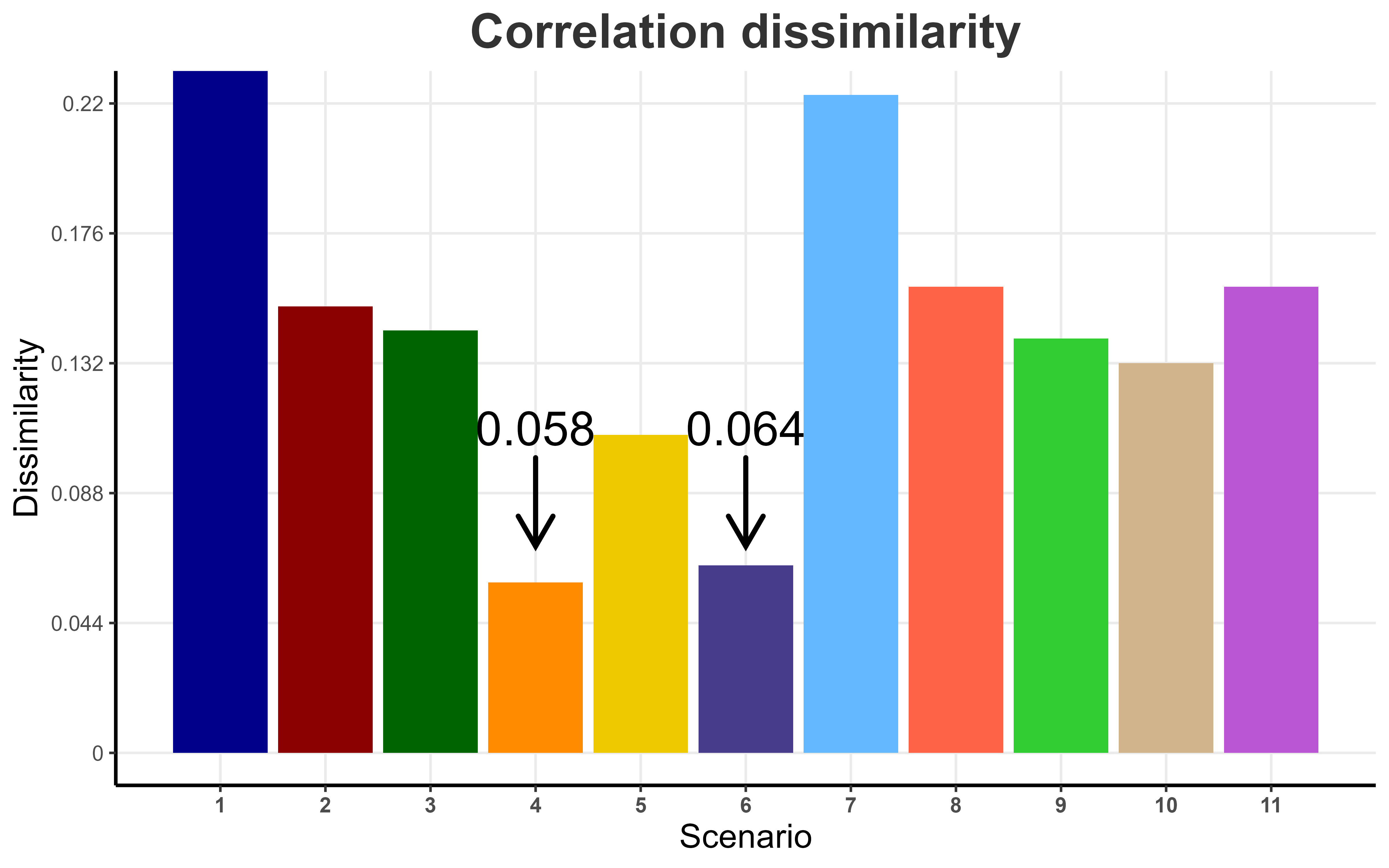}
\end{center}
\caption{The 11 method-1 scenarios tested by the correlation-based dissimilarity. The best two scenarios are number 4 (sample twice per week from 9 WWTP's) and number 6 (sample twice per week from 24 WWTP's).}
\label{fig:corsimmth1}
\end{figure}

Our conclusion is that, using the method-1 aggregated statistic, the most similar scenario to the reference curve is achieved by sampling twice per week from 9 WWTP's, while using the method-2 national curve, the best scenario is sampling once per week from 48 WWTP's. Taking into account the two best scenarios from each method, according to the method-1 statistic, we can reduce the number of treatment plants, as long as they cover the whole Austria and we should not reduce sampling frequency. According to the method-2 statistic, we can reduce frequency to once per week retaining 48 WWTP's, but if we choose to reduce the number of WWTP's to 24, then we cannot reduce sampling frequency. Therefore, finding some common ground between the two methods, we can say that we should not reduce sampling frequency, but it is safe to keep only 24 WWTP's. In Appendix E, we follow a similar procedure to evaluate the difference that sewer types and sizes make to the national curve. We find that using sewer systems combined with drain water of WWTP's of large size (over $10^5$ population size in the catchment area) provides the closest scenario with the original national curve.

\section{Statistical process monitoring}
\label{sec:spm}

\begin{figure}[h!]
\begin{center}
\includegraphics[width=0.9\textwidth]{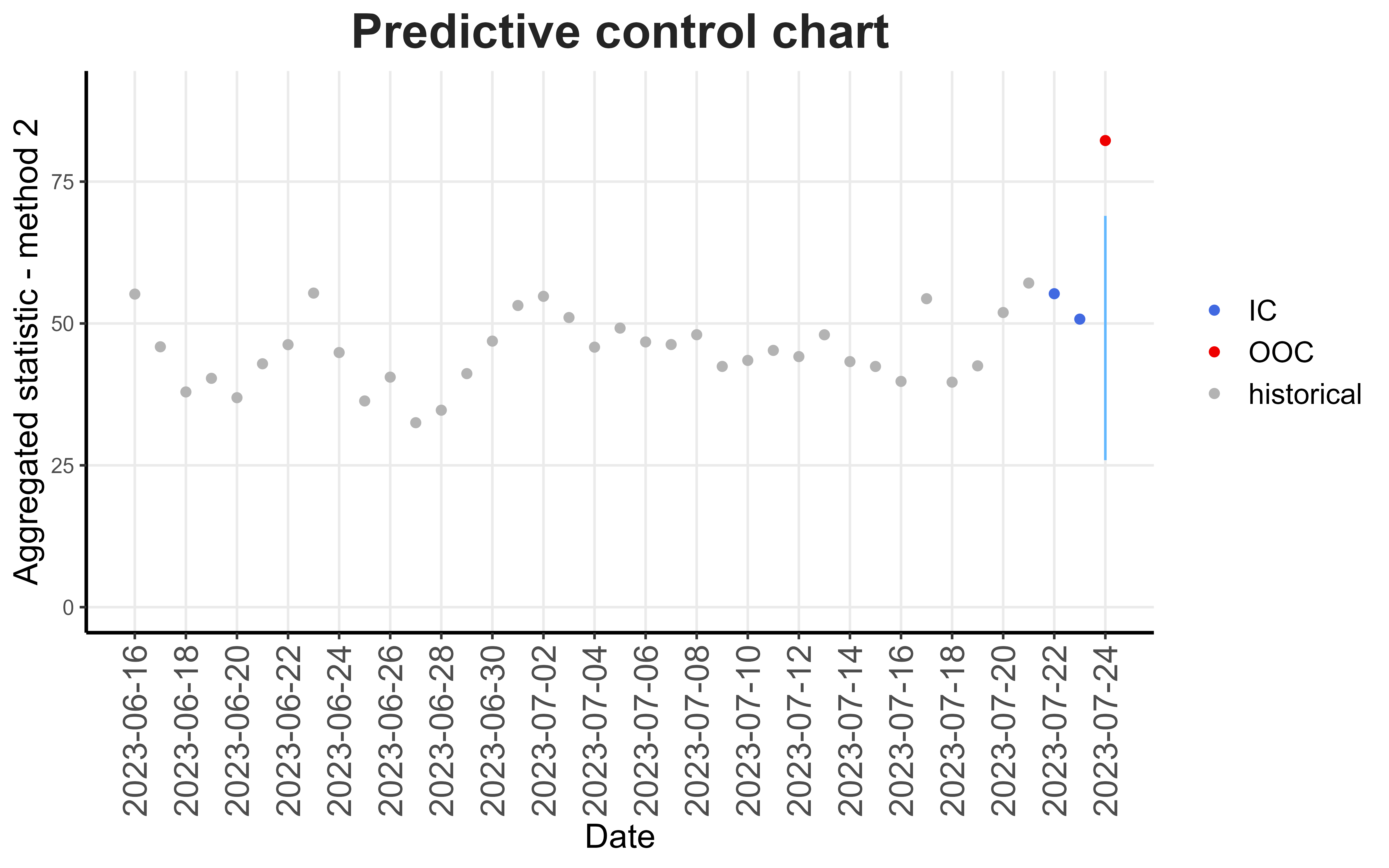}
\end{center}
\caption{The predictive control chart using only the in-control (IC) data from 22/07/2023 and 23/07/2023 for training and 16/06/2023-21/07/2023 as historical data. The HPD region for the next point is illustrated with a vertical line and the 24/07/2023 datum is correctly identified as out-of-control (OOC).}
\label{fig:pcc_meth2}
\end{figure}

Statistical process monitoring (SPM) is a procedure, where one wishes to locate the presence of a fault in a process and be warned by an alarm when this happens. When a fault occurs, it is the outcome of an assignable cause, which introduces excess noise in the process, which otherwise operates under only its natural variability. The tool to monitor the evolution of the process and produce an alarm, when it seems that assignable causes are present, is the control chart (for a comprehensive introduction to statistical process control, see \citealt{montgomery2020introduction}).

The majority of applications of SPM lies in the industry, where it is used to calibrate a process to operate under some desired specifications. However, the theory of SPM is not limited to such kinds of problems (see for instance \citealt{rogers2004control}) and can be naturally applied to a situation, where we want to keep track of the evolution of the national curve (our process) and receive an alarm, when the number of fictitious excretors is getting out of control.

The standard approach when a control chart is constructed is a two-phase experiment. During phase I, we calibrate the chart parameters when the process operates under no fault (it is in-control) and, during phase II we test the chart on new data. However, in our setting we cannot conduct such a phase I/II experiment, because we have already gathered the data and there is no clear in-control period to calibrate the chart. To this end, we only provide a proof-of-concept application of SPM, where we use an obvious in-control period as a pseudo phase I to calibrate a chart and then we test it sequentially to the data after that period.

As pseudo phase I, we consider the period 27/06/2023 until 19/07/2023 (for the method-1 statistic), or 16/06/2023 until 24/07/2023 (for the method-2 statistic), after visually inspecting the national curve and selecting the time interval with lowest values, which also coincides with a known decline of the virus spread. Then, phase II begins and the time of change corresponds to the first time point after the pseudo phase I. Thus, we expect the chart to generate an alarm immediately when phase II begins.

There are many different options regarding the control chart one can use and, for this work we test two families of charts: a Shewhart-type chart and a cumulative sum control chart (CUSUM). Generally we can say that the Shewhart $X$-chart works better with the method-2 statistic (because of less autocorrelation in the data, although we fix this issue with a modified version), while the CUSUM works better for the method-1 statistic. In Appendix F, we illustrate each chart's performance.

Another approach applicable when a phase I/II separation is impossible, due to limited or expensive data, or unethical in health-care cases, is on-line monitoring. Thus, even if an experiment with retrospective estimation of chart parameters is not possible in wastewater facilities, we propose the use of the predictive control chart of \citet{bourazas2022predictive}. This Bayesian chart plots the $100(1-\alpha)$\% highest predictive density (HPD) interval at each step and, if the newly obtained point lies outside this region, an alarm is generated. We test this approach on the method-2 aggregated statistic and we find it very promising. In Figure \ref{fig:pcc_meth2}, we plot the predictive control chart trained in \citealp{rstan} (instead of working analytically as \citealt{bourazas2022predictive} proposed), where we have used the previously used `pseudo phase I' data as historical data for prior determination. We assume that only the last two observations have been obtained and enable us to infer the next observation (correctly) as being out of control. The error rate $\alpha$ is determined so that it gives the same average run length (ARL) performance as the CUSUM chart.

\section{Model building}
\label{sec:modbuild}

Finally, since the aggregated statistics provide an estimation of the infected individuals in the population, we can use these data in a modeling approach. Thus, instead of cases-based epidemic models (like for instance \citealt{abbott2020estimating}), or deaths-based models (like \citealt{apsemidis2023bayesian}), the aggregated statistics can be seen as data for the total (both registered and unregistered) cases.

\begin{figure}[h!]
\begin{center}
\includegraphics[width=0.9\textwidth]{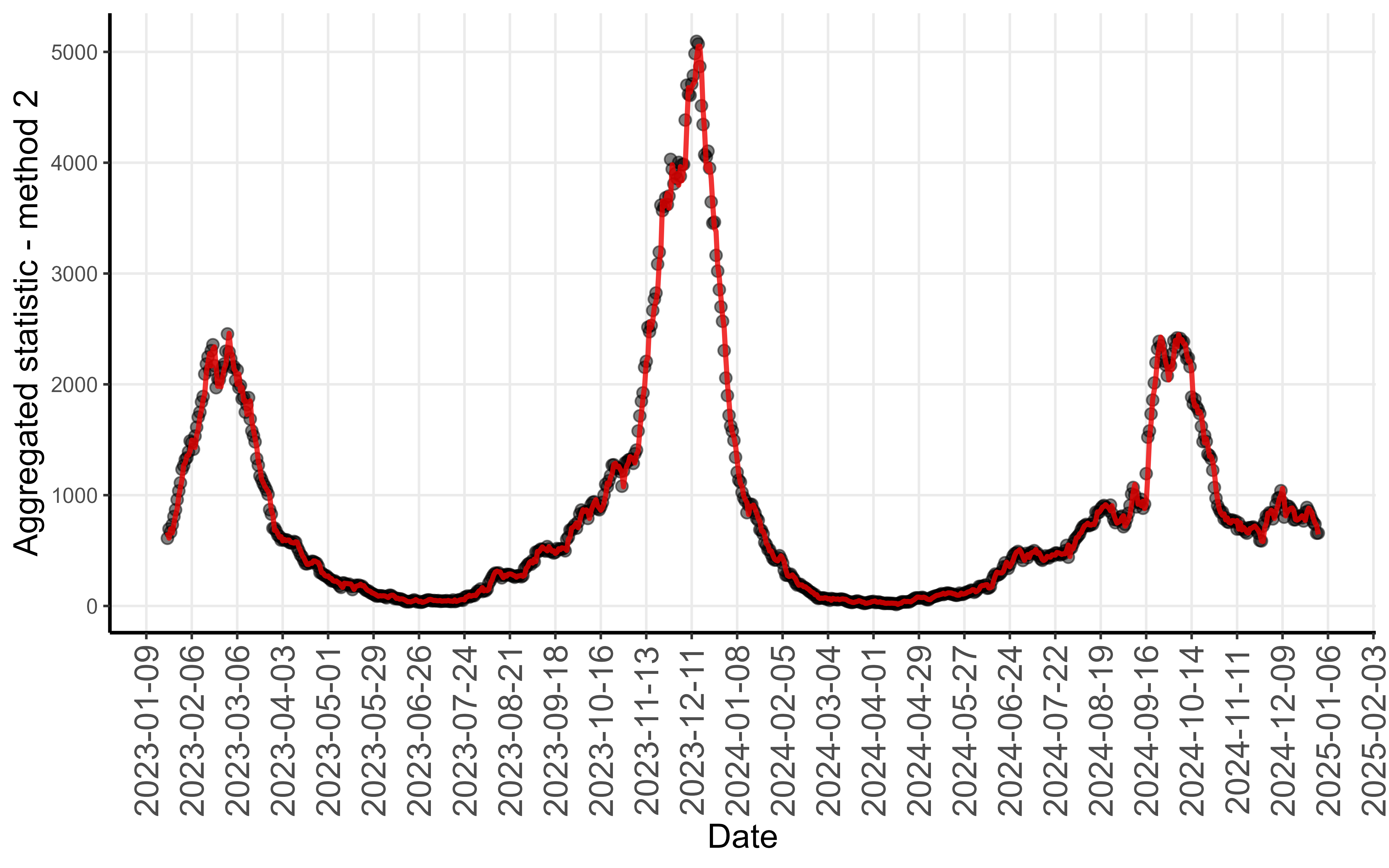}
\end{center}
\caption{Wastewater data of the method-2 aggregated statistic as grey points and fitted values of the count time series model as a red line.}
\label{fig:tscount}
\end{figure}

Another approach would be a time series model, like a seasonal ARIMA, or a count time series model (see for instance \citealt{fokianos2011log}), where we round the estimated infected individuals. As an illustration, we fit the following model using quasi conditional maximum likelihood
\begin{align*}
Y_t | \mathcal{F}_t &\sim NegBin(\lambda_t,395.25) \\
\lambda_t &= 0.037 + Y_{t-1} + 0.147\lambda_{t-2} + 0.012\lambda_{t-3} - 0.165\lambda_{t-4}
\end{align*}
where $Y_t$ is the rounded aggregated statistic (of method-2 in this instance), the parametrization of the Negative Binomial is in terms of mean and dispersion and $\mathcal{F}_t$ is the history of the process until time $t$. We train the model according to \citet{liboschik2017tscount} and the fitted values are shown in Figure \ref{fig:tscount}. This model is quite simple in the sense that it doesn't include any covariates, random effects, change-points, or any epidemic aspect (like for instance the infection rate) and, so we do not expect it to have any impressive prediction performance. However, it is a promising start and can be greatly enhanced, if the appropriate data are given.

\section{Discussion}
\label{sec:discuss}

WBE has gained much attention after the Covid-19 outbreak, where it was found to be very useful, since it allows access to lower levels of the epidemic pyramid (see Section \ref{sec:intro}). Motivated by a real application in Austria, we propose a statistical process control system that utilizes wastewater information and generates an alarm when the epidemic has gone out of control.

Specifically, we synthesize region-specific wastewater data to a national curve using two methods and provide estimates for the variability of each. These aggregated statistics are then used to evaluate the efficiency of the current sampling system answering the question: Is there any way to reduce costs without major loss of information? The results have shown that indeed, reducing the sampling sites from 48 to 24 keeping the same sampling frequency of twice per week is possible.

Then, we use this newly constructed aggregated statistics in a SPM scenario. We use two different SPM techniques: the phase I/II separation and on-line monitoring. Regarding the former, since we cannot conduct a proper experiment, we demonstrate the method using a `pseudo phase I' and CUSUM and Shewhart charts. If this approach seems preferable to the practitioner, then it should be performed after a phase I as described for instance in \citet{montgomery2020introduction} and \citet{chakraborti2008phase}. Regarding the on-line monitoring, we propose the use of the predictive control chart of \citet{bourazas2022predictive}, that requires no retrospective analysis and it can work immediately after the first two points become available. On the whole, we find the SPM idea fruitful and promising as a tool for WBE.

In the data, measurements on the chemical oxygen demand, nitrogen, and ammonium-nitrogen are included, which have been proven interesting in the literature. These may serve as covariates in the SPM setting to improve accuracy or on some trained prediction models, or as alternative methods of the estimated fictitious excretors, through the estimated grams per day that correspond to each infected person. We have not pursued such analysis in this article. Limitations of the control charts used are known in the SPM literature, like the effect of estimated parameters on the chart, or ARL performance (\citealt{jensen2006effects}). Finally, for the predictive control chart, when only trained using two data points (like we did in this application), keep in mind that the posterior is formed mainly by prior information, so the historically-informed prior is going to play a major role, until more data are obtained.

Some ideas for future research are other ways to aggregate the region-specific data to a national curve, taking into account other factors that we may have not used here. Moreover, we have assessed the variability of these statistics by sampling methods: either the natural extension provided by the method-2 statistic, or bootstrap based intervals. Analytical methods will improve upon this subject, since the data may not be enough for evaluating uncertainty for the method-2 statistic, or the time series model for parametric bootstrap may be wrong. Regarding SPM, other control charts may be more appropriate for such kinds of data (see for instance \citealt{apsemidis2020review}) and research on this subject can be very fruitful. Also, covariates (like the number of daily hospitalizations and the available beds in hospitals) can be inserted into the SPM context leading to regression charts, thus leading to a more complete support system. Finally, we illustrate a simple model on the aggregated national curve, but more flexible or complex models can be tested, like for example Gaussian processes (see \citealt{schulz2018tutorial}), splines (see \citealt{wood2017generalized}) or stochastic epidemic models (see \citealt{britton2010stochastic}).

\section*{Funding}

This research was done within the Austrian Agency for Health and Food Safety (AGES) task 5.5 of the KIRAS project `Wastewater surveillance as an instrument for crisis preparedness and pandemic management' funded by the Austrian Research Promotion Agency (FFG).

\bibliographystyle{apa}
\bibliography{bibtex}

\begin{thebibliography}{}

\bibitem[\protect\astroncite{Abbott et~al.}{2020}]{abbott2020estimating}
Abbott, S., Hellewell, J., Thompson, R. N., Sherratt, K., Gibbs, H. P., Bosse, N. I., Munday, J. D., Meakin, S., Doughty, E. L., Chun, J. Y., Finger, F., Campbell, P., CMMID COVID modelling group (2020).
\newblock Estimating the time-varying reproduction number of SARS-CoV-2 using national and subnational case counts.
\newblock {\em Wellcome Open Research}, 5(112):112.
\newblock \url{https://doi.org/10.12688/wellcomeopenres.16006.2}

\bibitem[\protect\astroncite{Apsemidis and Demiris}{2023}]{apsemidis2023bayesian}
Apsemidis, A. and Demiris, N. (2023).
\newblock Bayesian evidence synthesis for modeling SARS-CoV-2 transmission.
\newblock {\em arXiv preprint arXiv:2309.03122}.
\newblock \url{https://doi.org/10.48550/arXiv.2309.03122}

\bibitem[\protect\astroncite{Apsemidis et~al.}{2020}]{apsemidis2020review}
Apsemidis, A., Psarakis, S., and Moguerza, J. M. (2020).
\newblock A review of machine learning kernel methods in statistical process monitoring.
\newblock {\em Computers \& Industrial Engineering}, 142:106376.
\newblock \url{https://doi.org/10.1016/j.cie.2020.106376}

\bibitem[\protect\astroncite{Bertels et~al.}{2022}]{bertels2022factors}
Bertels, X., Demeyer, P., Van~den Bogaert, S., Boogaerts, T., van Nuijs, A. L., Delputte, P., and Lahousse, L. (2022).
\newblock Factors influencing SARS-CoV-2 RNA concentrations in wastewater up to the sampling stage: A systematic review.
\newblock {\em Science of the Total Environment}, 820:153290.
\newblock \url{https://doi.org/10.1016/j.scitotenv.2022.153290}

\bibitem[\protect\astroncite{Bourazas et~al.}{2022}]{bourazas2022predictive}
Bourazas, K., Kiagias, D., and Tsiamyrtzis, P. (2022).
\newblock Predictive Control Charts (PCC): A Bayesian approach in online monitoring of short runs.
\newblock {\em Journal of Quality Technology}, 54(4):367--391.
\newblock \url{https://doi.org/10.1080/00224065.2021.1916413}

\bibitem[\protect\astroncite{Britton}{2010}]{britton2010stochastic}
Britton, T. (2010).
\newblock Stochastic epidemic models: A survey.
\newblock {\em Mathematical biosciences}, 225(1):24--35.
\newblock \url{https://doi.org/10.1016/j.mbs.2010.01.006}

\bibitem[\protect\astroncite{Cevik et~al.}{2021}]{cevik2021sars}
Cevik, M., Tate, M., Lloyd, O., Maraolo, A.~E., Schafers, J., and Ho, A. (2021).
\newblock SARS-CoV-2, SARS-CoV, and MERS-CoV viral load dynamics, duration of viral shedding, and infectiousness: a systematic review and meta-analysis.
\newblock {\em The lancet microbe}, 2(1):e13--e22.
\newblock \url{https://doi.org/10.1016/S2666-5247(20)30172-5}

\bibitem[\protect\astroncite{Chakraborti et~al.}{2008}]{chakraborti2008phase}
Chakraborti, S., Human, S., and Graham, M. (2008).
\newblock Phase I Statistical Process Control Charts: An Overview and Some Results.
\newblock {\em Quality Engineering}, 21(1):52--62.
\newblock \url{https://doi.org/10.1080/08982110802445561}

\bibitem[\protect\astroncite{Ciannella et~al.}{2023}]{ciannella2023recent}
Ciannella, S., Gonz{\'a}lez-Fern{\'a}ndez, C., and Gomez-Pastora, J. (2023).
\newblock Recent progress on wastewater-based epidemiology for COVID-19 surveillance: A systematic review of analytical procedures and epidemiological modeling.
\newblock {\em Science of the Total Environment}, 878:162953.
\newblock \url{https://doi.org/10.1016/j.scitotenv.2023.162953}

\bibitem[\protect\astroncite{Fokianos and Tj{\o}stheim}{2011}]{fokianos2011log}
Fokianos, K. and Tj{\o}stheim, D. (2011).
\newblock Log-linear Poisson autoregression.
\newblock {\em Journal of multivariate analysis}, 102(3):563--578.
\newblock \url{https://doi.org/10.1016/j.jmva.2010.11.002}

\bibitem[\protect\astroncite{Gibbons et~al.}{2014}]{gibbons2014measuring}
Gibbons, C. L., Mangen, M.-J. J., Plass, D., Havelaar, A. H., Brooke, R. J., Kramarz, P., Peterson, K. L., Stuurman, A. L., Cassini, A., F{\`e}vre, E. M., and Kretzschmar, M. E.E. (2014).
\newblock Measuring underreporting and under-ascertainment in infectious disease datasets: a comparison of methods.
\newblock {\em BMC public health}, 14:1--17.
\newblock \url{https://doi.org/10.1186/1471-2458-14-147}

\bibitem[\protect\astroncite{Jensen et~al.}{2006}]{jensen2006effects}
Jensen, W. A., Jones-Farmer, L. A., Champ, C. W., and Woodall, W. H. (2006).
\newblock Effects of Parameter Estimation on Control Chart Properties: A Literature Review.
\newblock {\em Journal of Quality technology}, 38(4):349--364.
\newblock \url{https://doi.org/10.1080/00224065.2006.11918623}

\bibitem[\protect\astroncite{Kostoglou et~al.}{2024}]{kostoglou2024effect}
Kostoglou, M., Petala, M., Karapantsios, T., Dovas, C., Tsiridis, V., Roilides, E., Koutsolioutsou-Benaki, A., Paraskevis, D., Metalidis, S., Stylianidis, E., Papa, A., Papadopoulos, A., Tsiodras, S., and Papaioannou, N. (2024).
\newblock Effect of SARS-CoV-2 shedding rate distribution of individuals during their disease days on the estimation of the number of infected people. Application of wastewater-based epidemiology to the city of Thessaloniki, Greece.
\newblock {\em Science of the Total Environment}, 951:175724.
\newblock \url{https://doi.org/10.1016/j.scitotenv.2024.175724}

\bibitem[\protect\astroncite{Koureas et~al.}{2021}]{koureas2021wastewater}
Koureas, M., Amoutzias, G. D., Vontas, A., Kyritsi, M., Pinaka, O., Papakonstantinou, A., Dadouli, K., Hatzinikou, M., Koutsolioutsou, A., Mouchtouri, V. A., Speletas, M., Tsiodras, S., and Hadjichristodoulou, C. (2021).
\newblock Wastewater monitoring as a supplementary surveillance tool for capturing SARS-COV-2 community spread. A case study in two Greek municipalities.
\newblock {\em Environmental research}, 200:111749.
\newblock \url{https://doi.org/10.1016/j.envres.2021.111749}

\bibitem[\protect\astroncite{Liboschik et~al.}{2017}]{liboschik2017tscount}
Liboschik, T., Fokianos, K., and Fried, R. (2017).
\newblock tscount: An R Package for Analysis of Count Time Series Following Generalized Linear Models.
\newblock {\em Journal of Statistical Software}, 82:1--51.
\newblock \url{https://doi.org/10.18637/jss.v082.i05}

\bibitem[\protect\astroncite{Medema et~al.}{2020}]{medema2020implementation}
Medema, G., Been, F., Heijnen, L., and Petterson, S. (2020).
\newblock Implementation of environmental surveillance for SARS-CoV-2 virus to support public health decisions: Opportunities and challenges.
\newblock {\em Current opinion in environmental science \& health}, 17:49--71.
\newblock \url{https://doi.org/10.1016/j.coesh.2020.09.006}

\bibitem[\protect\astroncite{Miura et~al.}{2021}]{miura2021duration}
Miura, F., Kitajima, M., and Omori, R. (2021).
\newblock Duration of SARS-CoV-2 viral shedding in faeces as a parameter for wastewater-based epidemiology: Re-analysis of patient data using a shedding dynamics model.
\newblock {\em Science of The Total Environment}, 769:144549.
\newblock \url{https://doi.org/10.1016/j.scitotenv.2020.144549}

\bibitem[\protect\astroncite{Montgomery}{2020}]{montgomery2020introduction}
Montgomery, D. C. (2020).
\newblock {\em Introduction to statistical quality control}.
\newblock John wiley \& sons.

\bibitem[\protect\astroncite{P{\'a}jaro et~al.}{2022}]{pajaro2022stochastic}
P{\'a}jaro, M., Fajar, N. M., Alonso, A. A., and Otero-Muras, I. (2022).
\newblock Stochastic SIR model predicts the evolution of COVID-19 epidemics from public health and wastewater data in small and medium-sized municipalities: A one year study.
\newblock {\em Chaos, Solitons \& Fractals}, 164:112671.
\newblock \url{https://doi.org/10.1016/j.chaos.2022.112671}

\bibitem[\protect\astroncite{Rauch et~al.}{2024}]{rauch2024estimating}
Rauch, W., Schenk, H., Rauch, N., Harders, M., Oberacher, H., Insam, H., Markt, R., and Kreuzinger, N. (2024).
\newblock Estimating actual SARS-CoV-2 infections from secondary data.
\newblock {\em Scientific Reports}, 14(1):6732.
\newblock \url{https://doi.org/10.1038/s41598-024-57238-0}

\bibitem[\protect\astroncite{Riedmann et~al.}{2025}]{riedmann2025estimates}
Riedmann, U., Chalupka, A., Richter, L., Sprenger, M., Rauch, W., Schenk, H., Krause, R., Willeit, P., Oberacher, H., H{\o}eg, T. B., Ioannidis, J. P. A., and Pilz, S. (2025).
\newblock Estimates of SARS-CoV-2 infections and population immunity after the COVID-19 pandemic in Austria: Analysis of national wastewater data.
\newblock {\em The Journal of Infectious Diseases}, page jiaf054.
\newblock \url{https://doi.org/10.1093/infdis/jiaf054}

\bibitem[\protect\astroncite{Rogers et~al.}{2004}]{rogers2004control}
Rogers, C. A., Reeves, B. C., Caputo, M., Ganesh, J. S., Bonser, R. S., and Angelini, G. D. (2004).
\newblock Control chart methods for monitoring cardiac surgical performance and their interpretation.
\newblock {\em The Journal of Thoracic and Cardiovascular Surgery}, 128(6):811--819.
\newblock \url{https://doi.org/10.1016/j.jtcvs.2004.03.011}

\bibitem[\protect\astroncite{Rose et~al.}{2015}]{rose2015characterization}
Rose, C., Parker, A., Jefferson, B., and Cartmell, E. (2015).
\newblock The Characterization of Feces and Urine: A Review of the Literature to Inform Advanced Treatment Technology.
\newblock {\em Critical reviews in environmental science and technology},45(17):1827--1879.
\newblock \url{https://doi.org/10.1080/10643389.2014.1000761}

\bibitem[\protect\astroncite{Schulz et~al.}{2018}]{schulz2018tutorial}
Schulz, E., Speekenbrink, M., and Krause, A. (2018).
\newblock A tutorial on Gaussian process regression: Modelling, exploring, and exploiting functions
\newblock {\em Journal of mathematical psychology}, 85:1--16.
\newblock \url{https://doi.org/10.1016/j.jmp.2018.03.001}

\bibitem[\protect\astroncite{Somerset and Brown}{2024}]{somerset2024wastewater}
Somerset, E. and Brown, P. E. (2024).
\newblock Wastewater surveillance using differentiable Gaussian processes.
\newblock {\em Journal of the Royal Statistical Society Series C: Applied Statistics}, page qlae073.
\newblock \url{https://doi.org/10.1093/jrsssc/qlae073}

\bibitem[\protect\citeauthoryear{{Stan}}{}{}]{rstan}
Stan Development Team (2025).
\newblock {RStan}: the {R} interface to {Stan}.
\newblock R package version 2.32.7.

\bibitem[\protect\astroncite{Teunis et~al.}{2015}]{teunis2015shedding}
Teunis, P., Sukhrie, F., Vennema, H., Bogerman, J., Beersma, M., and Koopmans, M. (2015).
\newblock Shedding of norovirus in symptomatic and asymptomatic infections.
\newblock {\em Epidemiology \& Infection}, 143(8):1710--1717.
\newblock \url{doi:10.1017/S095026881400274X}

\bibitem[\protect\astroncite{Wood}{2017}]{wood2017generalized}
Wood, S. N. (2017).
\newblock Generalized additive models: an introduction with R.
\newblock {\em Chapman and Hall/CRC}.
\newblock \url{https://doi.org/10.1201/9781315370279}

\bibitem[\protect\astroncite{Zhang et~al.}{2022}]{zhang2022sars}
Zhang, D., Duran, S. S. F., Lim, W. Y. S., Tan, C. K. I., Cheong, W. C. D., Suwardi, A., and Loh, X. J. (2022).
\newblock SARS-CoV-2 in wastewater: From detection to evaluation.
\newblock {\em Materials Today Advances}, 13:100211.
\newblock \url{https://doi.org/10.1016/j.mtadv.2022.100211}

\end{thebibliography}


\begin{thebibliography}{}

\bibitem[\protect\astroncite{DiCiccio and Efron}{1996}]{diciccio1996bootstrap}
DiCiccio, T. J. and Efron, B. (1996).
\newblock Bootstrap confidence intervals.
\newblock {\em Statistical science}, 11(3):189--228.
\newblock \url{https://doi.org/10.1214/ss/1032280214}

\bibitem[\protect\astroncite{Efron}{1981}]{efron1981nonparametric}
Efron, B. (1981).
\newblock Nonparametric standard errors and confidence intervals.
\newblock {\em canadian Journal of Statistics}, 9(2):139--158.
\newblock \url{https://doi.org/10.2307/3314608}

\end{thebibliography}

\end{document}


\def\spacingset#1{\renewcommand{\baselinestretch}%
{#1}\small\normalsize} \spacingset{1}

\thispagestyle{plain}
\begin{center}
\Large{\textbf{Supplementary Material for Integrating Region-Specific SARS-CoV-2 Data for Statistical Wastewater Monitoring}}
\end{center}

\section*{Appendix A. Austrian dataset of wastewater measurements}

In table \ref{tab:dataset}, the Austrian dataset is shown. It comprises of 13 variables for 48 WWTP's over 9 states, which have to be synthesized for a nation-wide monitoring system.

\begin{table}[!h]
\caption{The initial form of the wastewater data.}
\label{tab:dataset}
\begin{center}
\begin{tabular}{ c l } 
\hline
Variable & Description \\ \hline
1 & anonymized key for wastewater treatment plant (48 keys) \\
2 & if WWTP was monitored fom the beginning or was added later (2 levels) \\
3 & federal state (9 levels) \\
4 & type of sewer system (3 types) \\
5 & date of sampling (date) \\
6 & sample ID from laboratory \\
7 & virus concentration (numeric) \\
8 & inflow of wastewater on the sampling day in the sewer (numeric) \\
9 & inflow temperature (numeric) \\
10 & residents in the catchment area of the WWTP (numeric) \\ 
11 & chemical oxygen demand (numeric) \\
12 & total nitrogen (numeric) \\ 
13 & ammonium-nitrogen (numeric) \\ 
\hline
\end{tabular}
\end{center}
\end{table}

\section*{Appendix B. Aggregation methods for the national curve}

\begin{figure}[h!]
\begin{center}
\includegraphics[width=0.9\textwidth]{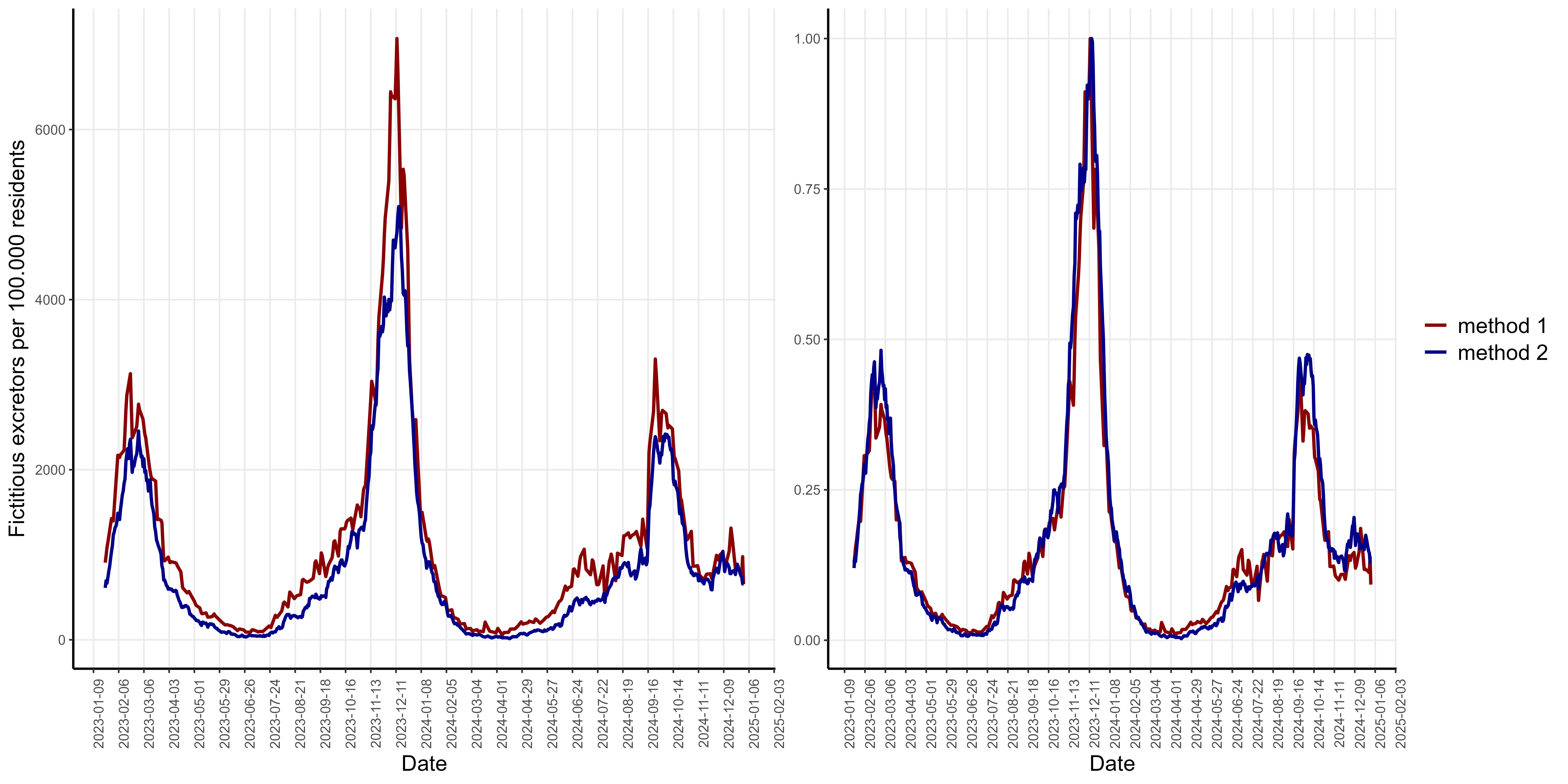}
\end{center}
\caption{Left: The two national curves on the same plot. Right: The normalized version of the two curves, where we have divided by the maximum of its, to rescale them inside the $(0,1)$ interval. In both versions, the two methods seem to be very similar.}
\label{fig:methods_12t}
\end{figure}

\begin{figure}[h!]
\begin{center}
\includegraphics[width=0.9\textwidth]{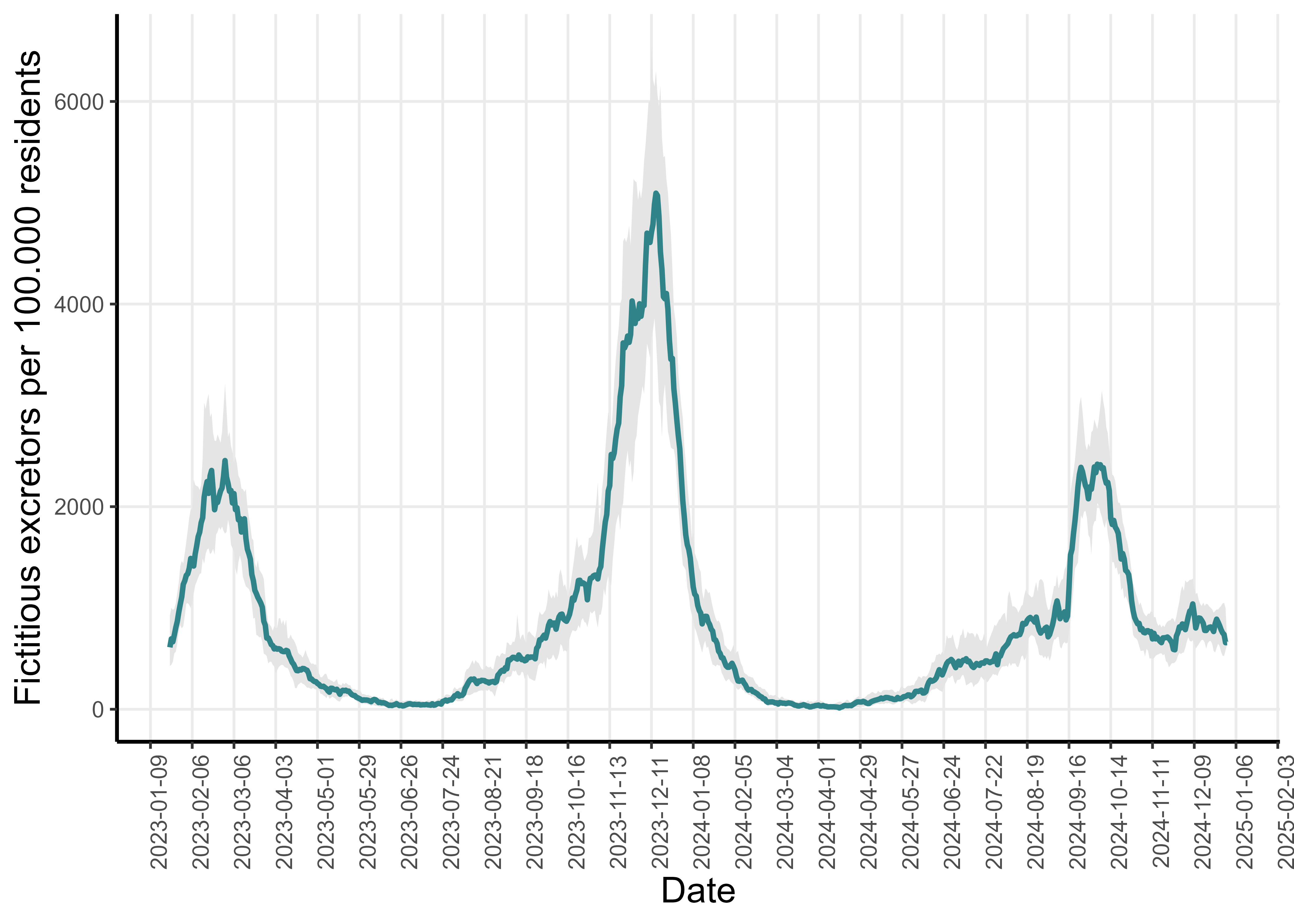}
\end{center}
\caption{A natural direct way of assessing the variability of the method-2 statistic is the interquartile range, depicted here as grey ribbon around the green median.}
\label{fig:ficex_aggr2}
\end{figure}

In Figure \ref{fig:methods_12t}, the national curves under the two aggregation methods are shown. In order to put them on the same scale, we further normalize them into the $(0,1)$ interval by dividing each value with the maximum of the corresponding curve. Also, in Figure \ref{fig:ficex_aggr2}, we provide an easy way to assess the variability of the method-2 statistic, by using the difference of the 0.75 and 0.25 percentiles (i.e. the interquartile range) at each day.

\section*{Appendix C. Assessing variability using bootstrap}

Herein, the inherent variability of each statistic is considered. For the method-2 statistic, we can directly assess its variance by the available samples of each day. Thus, a point-wise percentile confidence interval can be constructed by
$$(\kappa_{\alpha/2},\kappa_{1-\alpha/2})$$
where the $\kappa$ values denote the percentiles at values defined via the confidence level $\alpha$. 

\begin{figure}[h!]
\begin{center}
\includegraphics[width=0.9\textwidth]{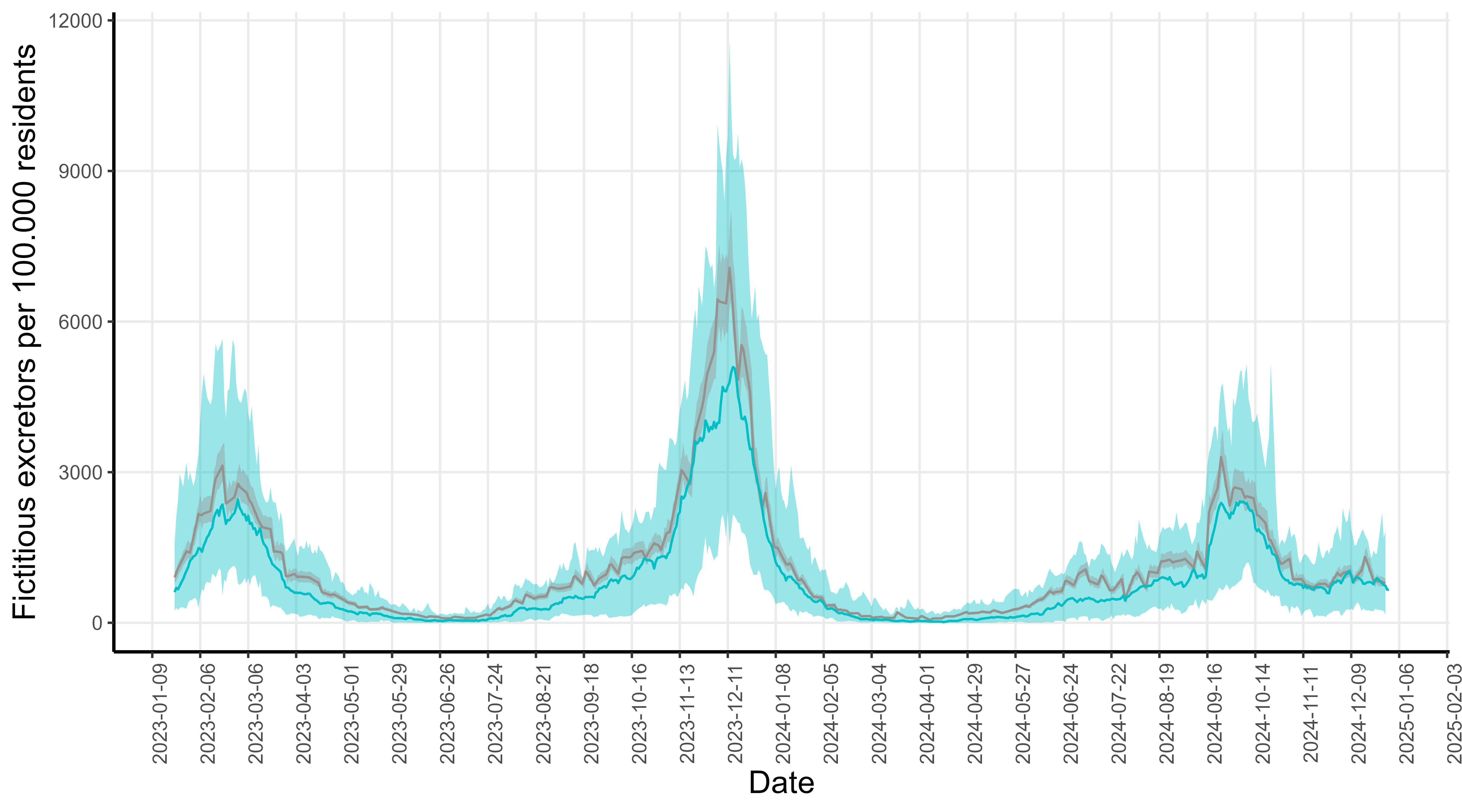}
\end{center}
\caption{The two national curves on the same plot along with 95\% point-wise confidence intervals. In red is the method-1 statistic and in light blue is the method-2 statistic.}
\label{fig:natcuinter}
\end{figure}

On the other hand, the method-2 statistic has no obvious standard error, due to the unknown density of the concentration of the virus in the water. Thus, one solution is to use bootstrap and apply the percentile interval on the gathered bootstrap values (see \citealt{diciccio1996bootstrap} and \citealt{efron1981nonparametric}; (the algorithm is depicted in algorithm \ref{alg:bootstrapci}). The national curves are shown in Figure \ref{fig:natcuinter} along with point-wise 95\% confidence intervals. The method-2 statistic has wider intervals, since they rely on much fewer samples (the 48 WWTP measurements per day), while the credibility of the method-1 intervals relies on the assumption that the parametric time series model used for bootstrap holds true. Note that the two methods are not on the same scale for a fair comparison.

\begin{algorithm}[h!]
\caption{Bootstrap Percentile CI for time series}\label{alg:bootstrapci}
\SetAlgoLined
\KwInput{Fitted values $\hat{y}_{1:T}$, residuals $r_{1:T}$, number of bootstrap replications $B$}
Set $j=1$ \\
\Repeat{$j=B$}{
Choose a sample of size $T$ from the residuals $r_1^{(b)},...,r_T^{(b)}$ \\
Calculate a new time series $y_{1:T}^{(b)}=y_{1:T}+r_{1:T}^{(b)}$ \\
Set $j=j+1$}
\KwOutput{Return quantiles of $y_t^{(b)}$ for each day $t$}
\end{algorithm}

\section*{Appendix D. Influence of WWTP's on the national curve}

\begin{figure}[h!]
\begin{center}
\includegraphics[width=0.9\textwidth]{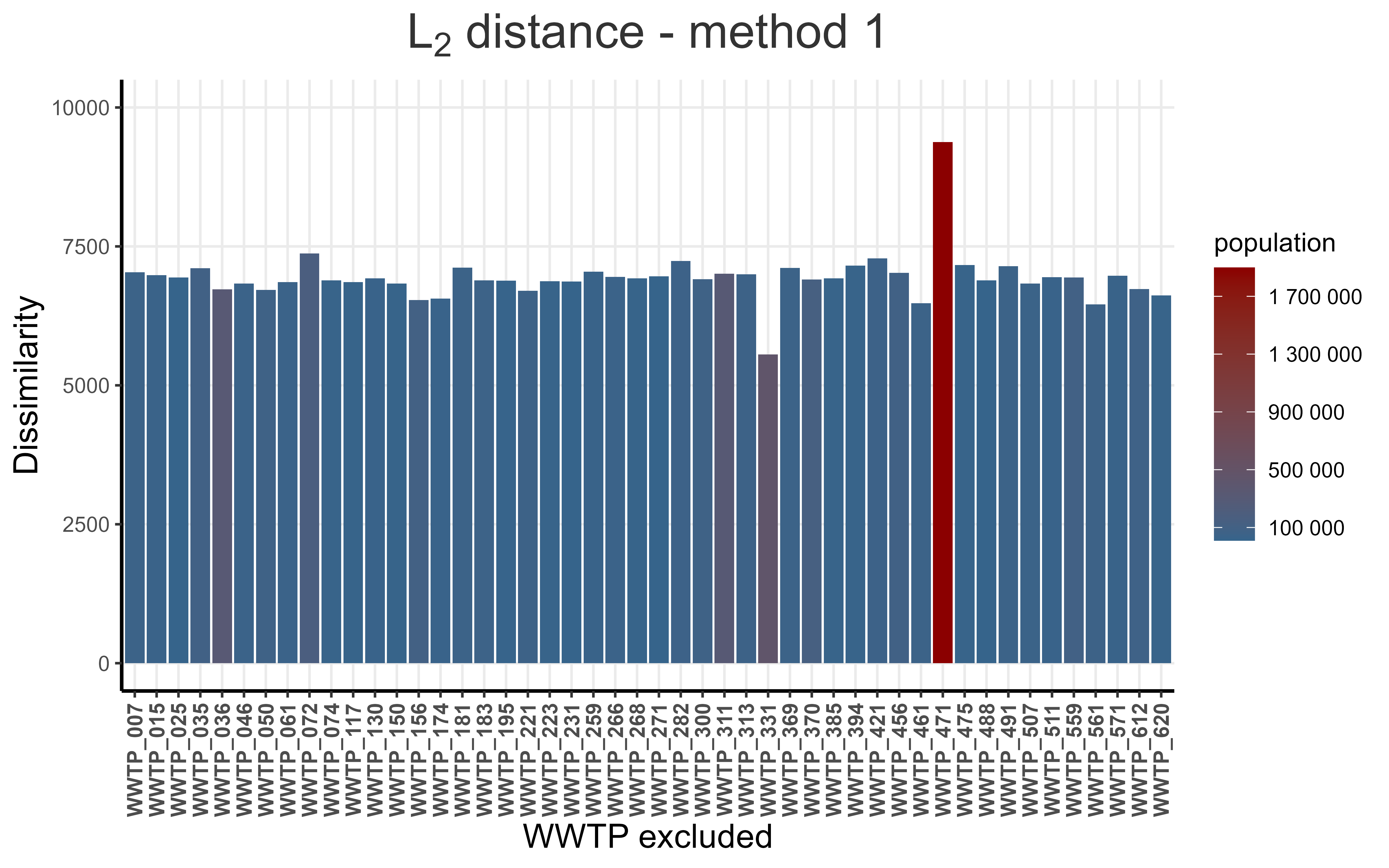}
\includegraphics[width=0.9\textwidth]{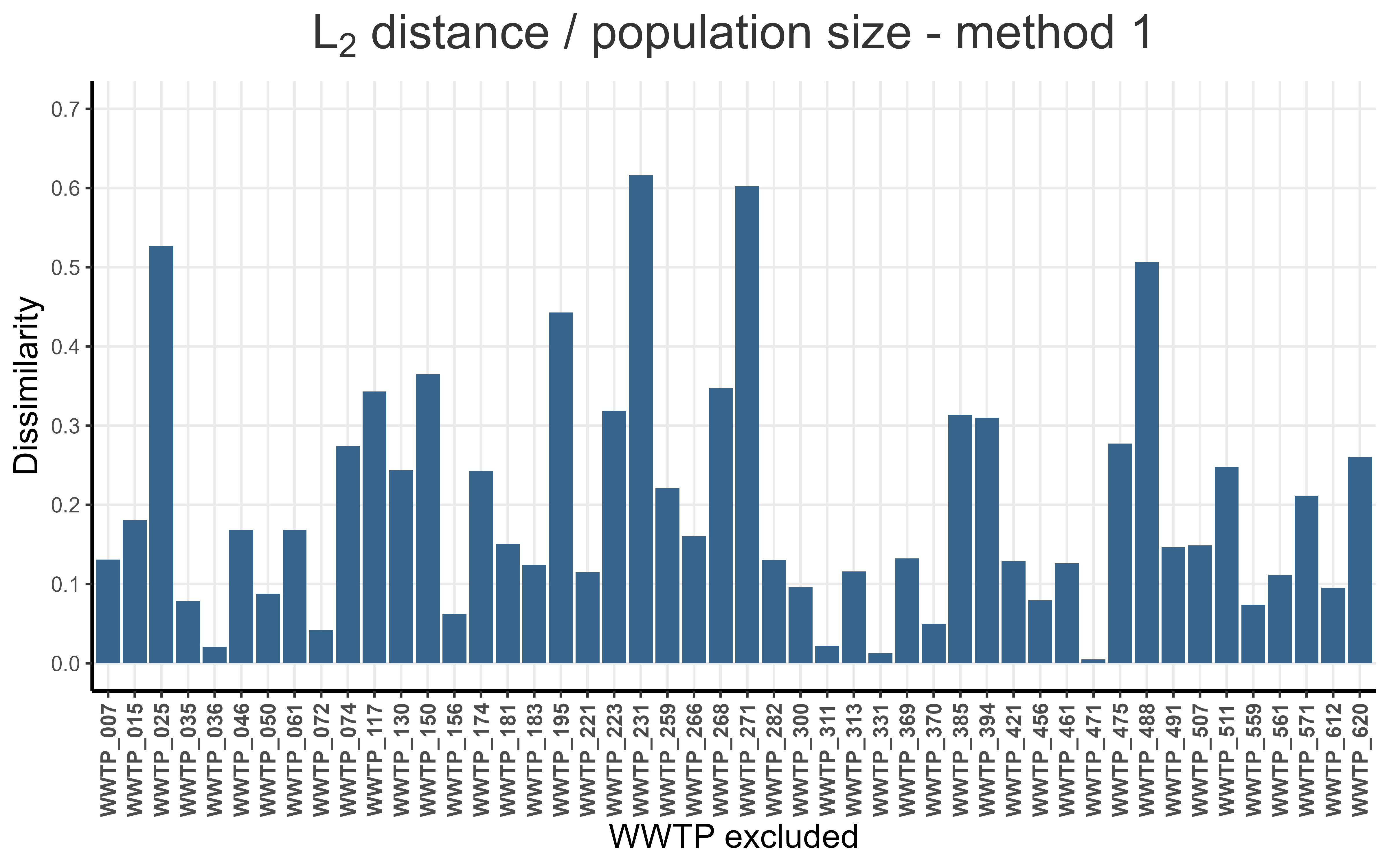}
\end{center}
\caption{Influence of each WWTP on the method-1 national curve. Left: Without taking into account population size. Right: Normalized by population size.}
\label{fig:inflmeth1}
\end{figure}

Of course, not every WWTP plays the same role in the determination of the national curve and so we proceed quantifying a WWTP's individual influence, defined as its ability to change the curve, when we remove it from the aggregation method. The algorithm is depicted in \ref{alg:influenceal}, while the results for the method-1 statistic are shown on the upper side of Figure \ref{fig:inflmeth1} (the rest lie in Appendix D). Since, a major factor in the influence estimation is the population size in the catchment area, we also provide a normalized version on the lower side of Figure \ref{fig:inflmeth1}, where we divide its influence by the corresponding population size. Thus, we can say that the most influential WWTP's are two from Lower Austria and one from Styria (if population size is not taken into account, then Vienna takes over). 

\begin{algorithm}[h!]
\caption{WWTP influence estimation}\label{alg:influenceal}
\SetAlgoLined
\KwInput{Wastewater data $x_{1:T,1:J}$, aggregated data $y_{1:T}$ for $T$ days and $J$ WWTP's}
\For{WWTP index $j=1,...,J$}{
Remove rows referring to WWTP $j$, $x_{1:T,-j}$ \\
Aggregate data to the national level, $y_{1:T,-j}$ \\
Calculate the dissimilarity $D_j(\cdot,\cdot)$ with the original time series, $D_j(y_{1:T},y_{1:T,-j})$}
\KwOutput{Return $J$ dissimilarities}
\end{algorithm}
For the bootstrap confidence interval (CI), first we need to fit a time series model on the national curve $y_{1:T}$ and obtain fitted values $\hat{y}_{1:T}$ and residuals $r_{1:T}$. Then, we follow Algorithm \ref{alg:bootstrapci}.

\begin{figure}[h!]
\begin{center}
\includegraphics[width=0.9\textwidth]{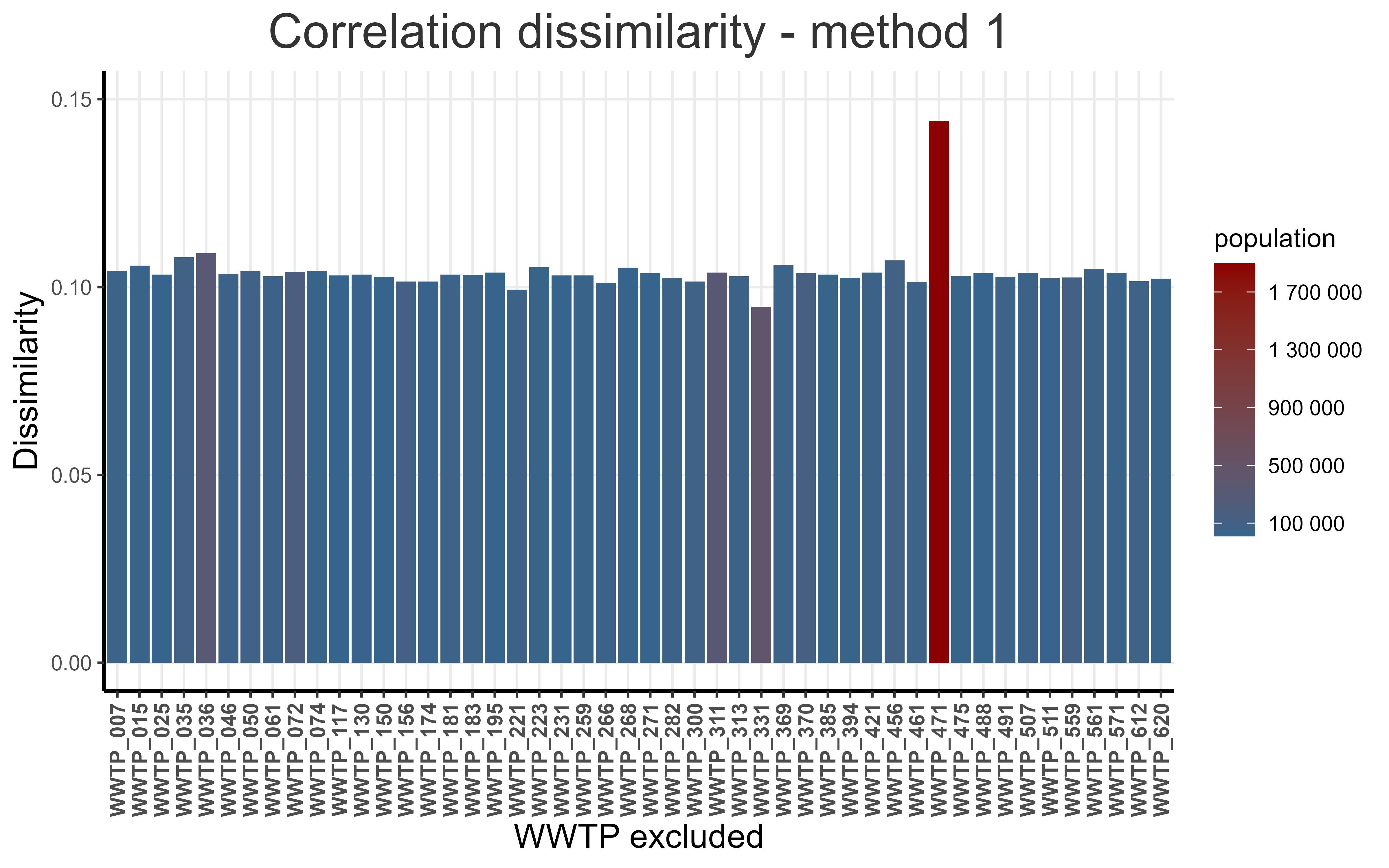}
\includegraphics[width=0.9\textwidth]{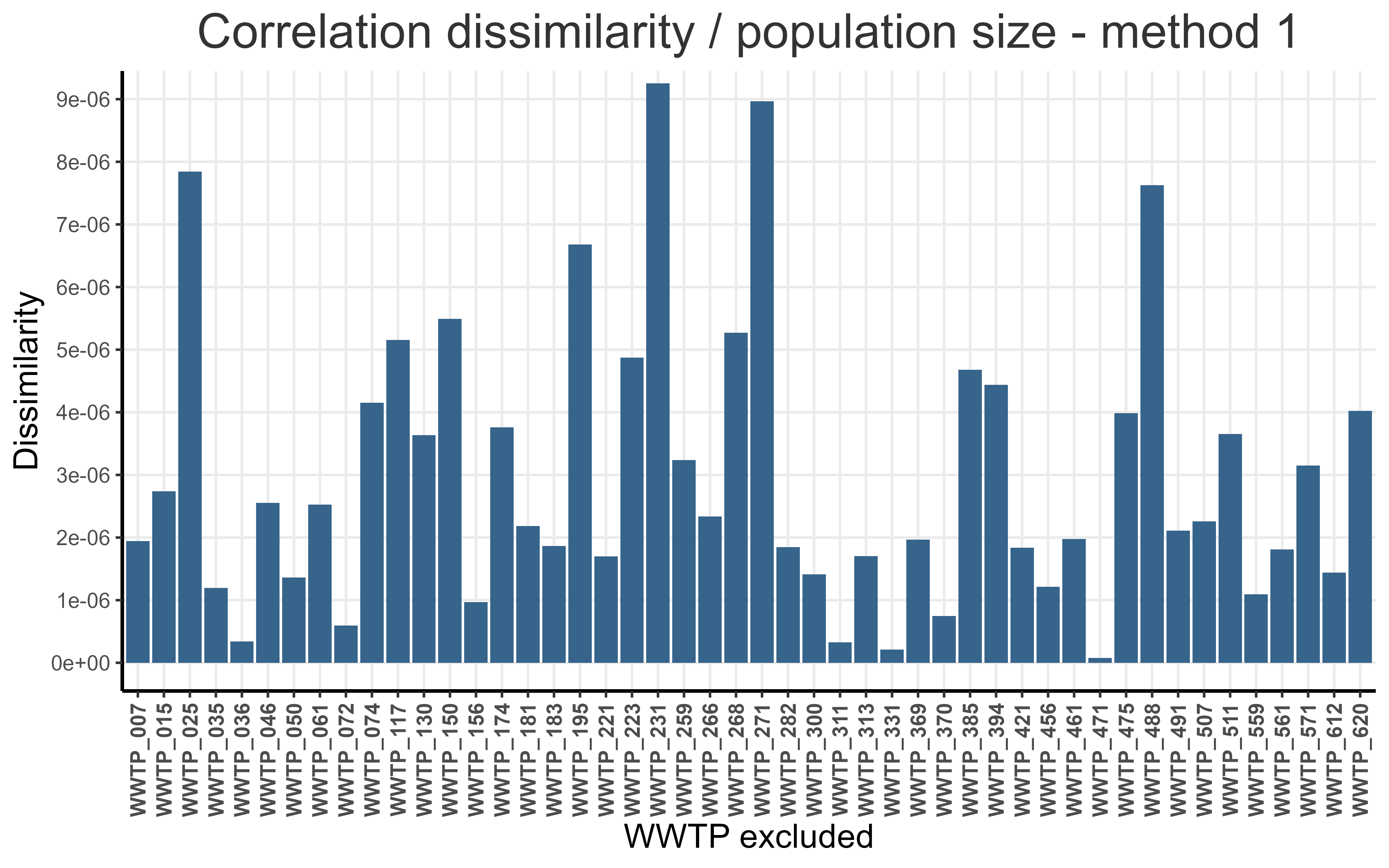}
\end{center}
\caption{Influence of each WWTP on the method-1 national curve. Left: Without taking into account population size. Right: Normalized by population size.}
\label{fig:cordispop_meth1}
\end{figure}

\begin{figure}[h!]
\begin{center}
\includegraphics[width=0.9\textwidth]{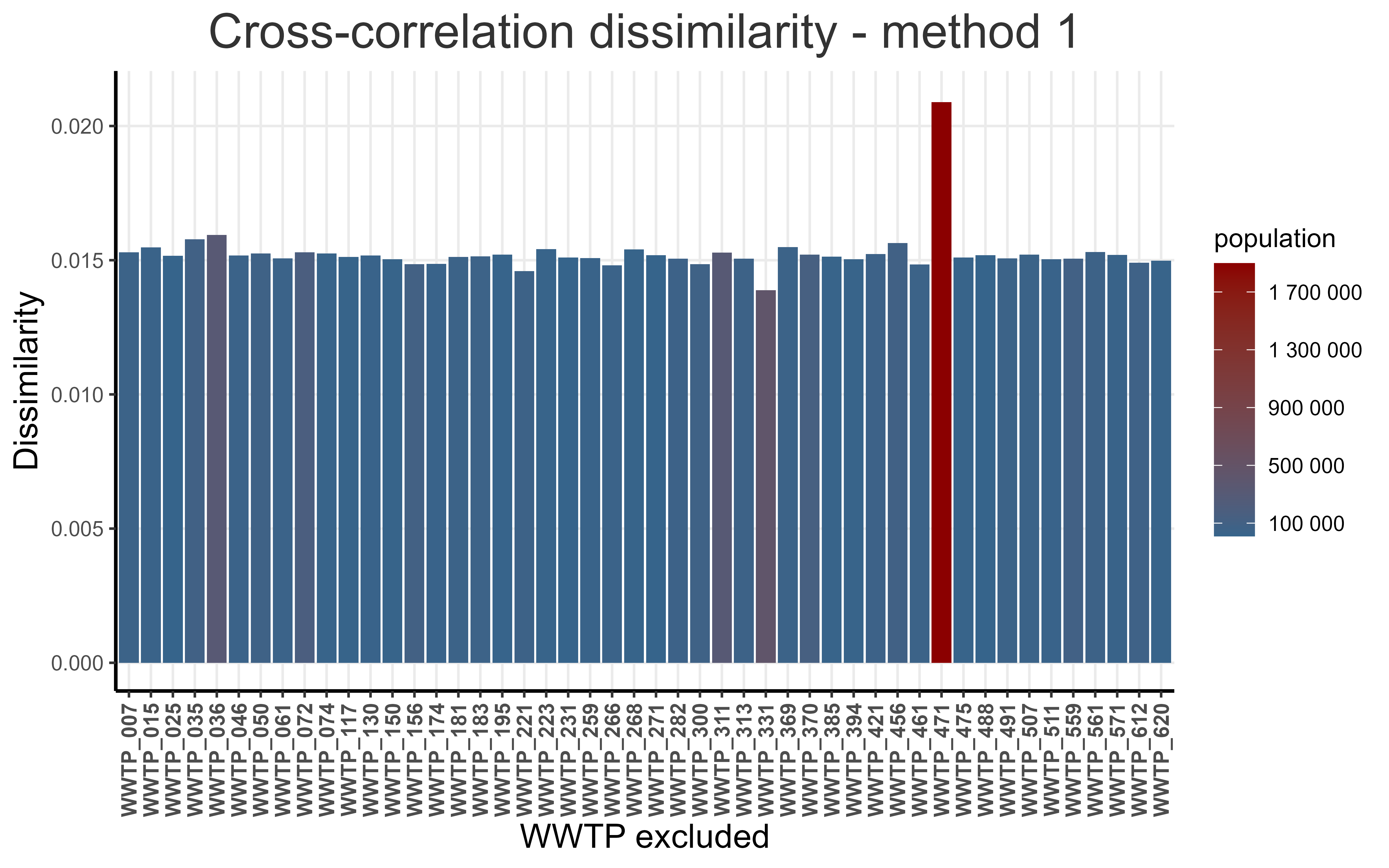}
\includegraphics[width=0.9\textwidth]{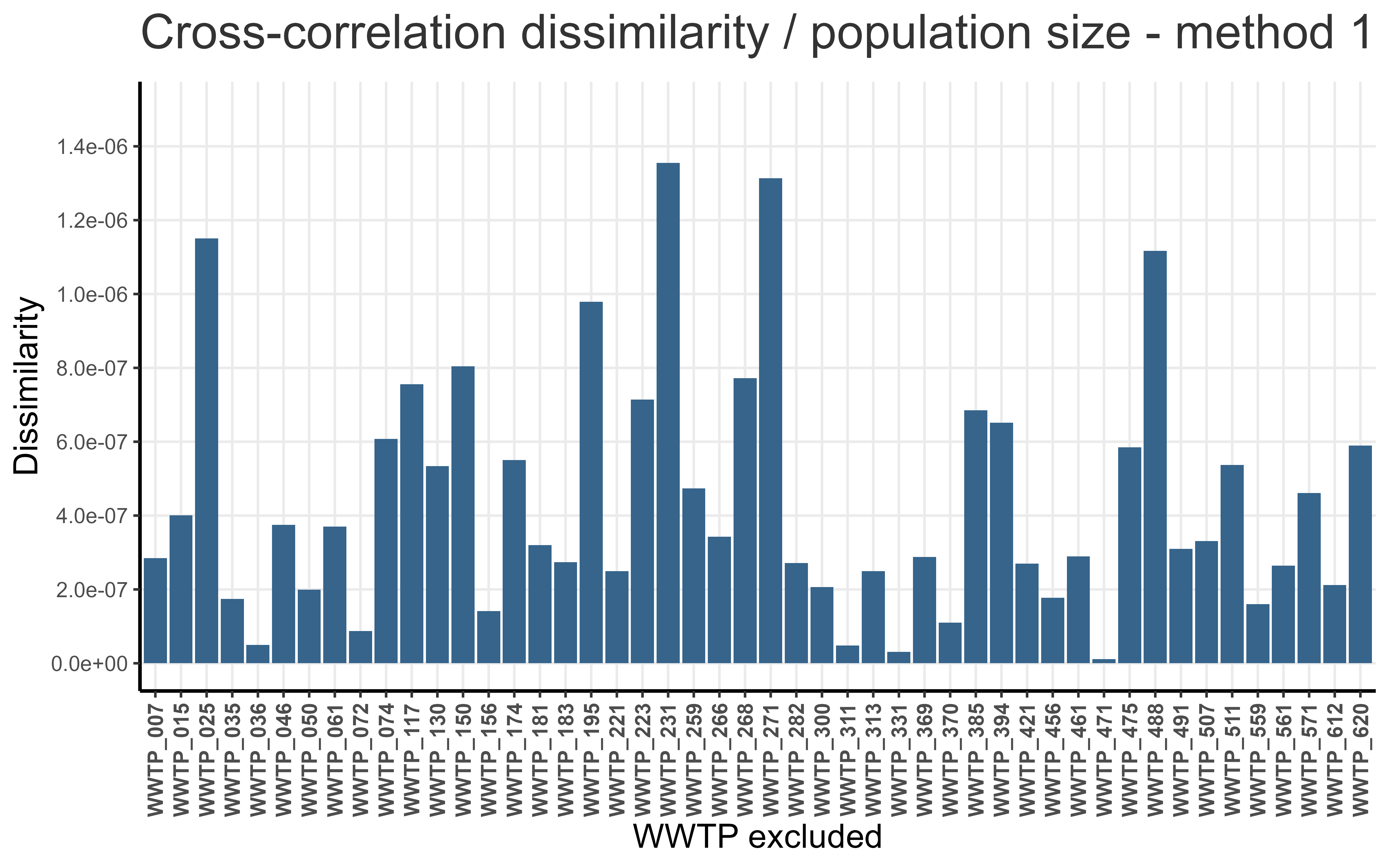}
\end{center}
\caption{Influence of each WWTP on the method-1 national curve. Left: Without taking into account population size. Right: Normalized by population size.}
\label{fig:ccordispop_meth1}
\end{figure}

Below we show the results using the correlation (Figure \ref{fig:cordispop_meth1}) and cross-correlation (Figure \ref{fig:ccordispop_meth1}) measures. Since the method-2 statistic mixes the available samples of each day, we cannot use the same technique for it.

\section*{Appendix E. Investigation of sub-sampling scenarios}

\begin{figure}[h!]
\begin{center}
\includegraphics[width=0.9\textwidth]{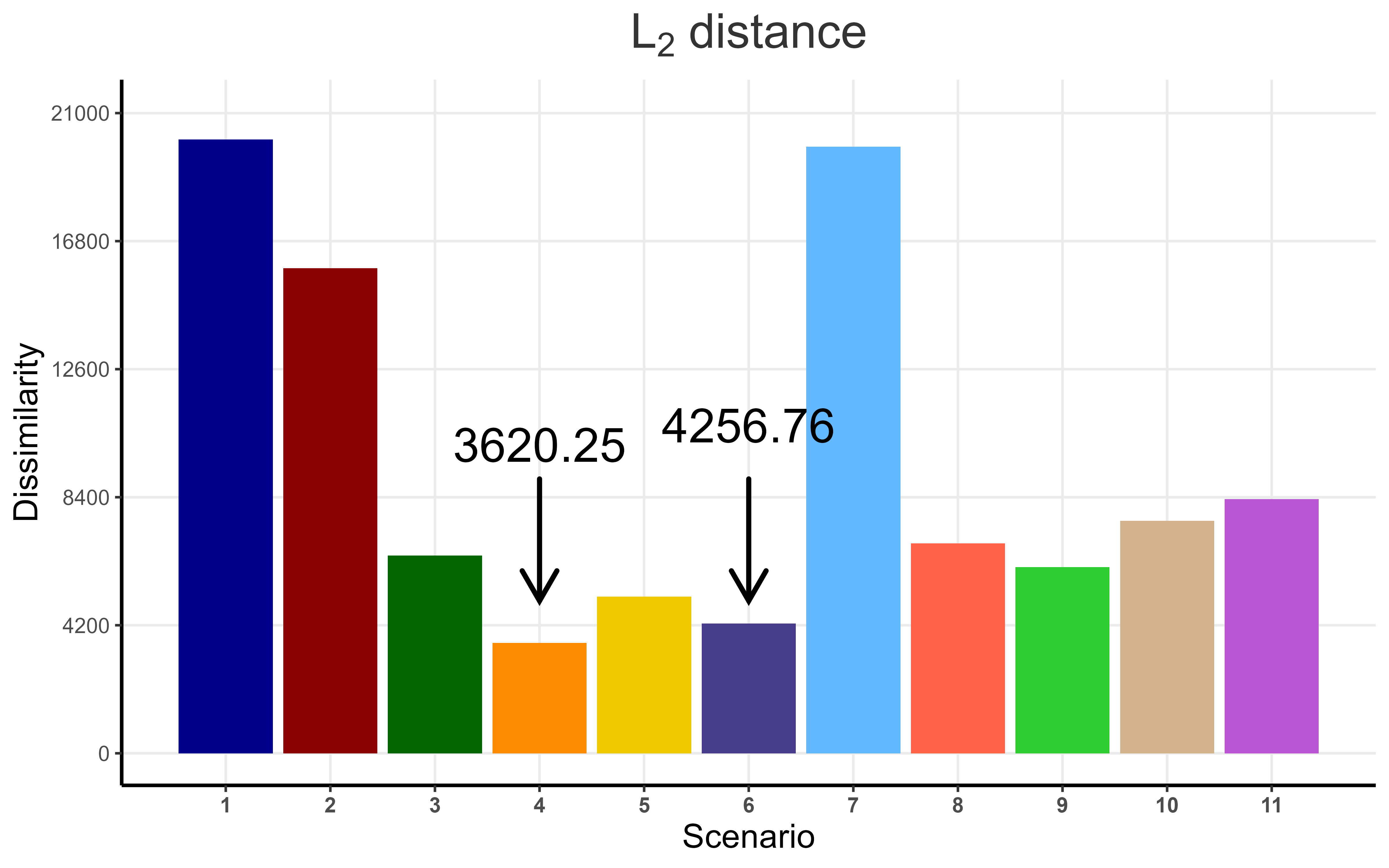}
\includegraphics[width=0.9\textwidth]{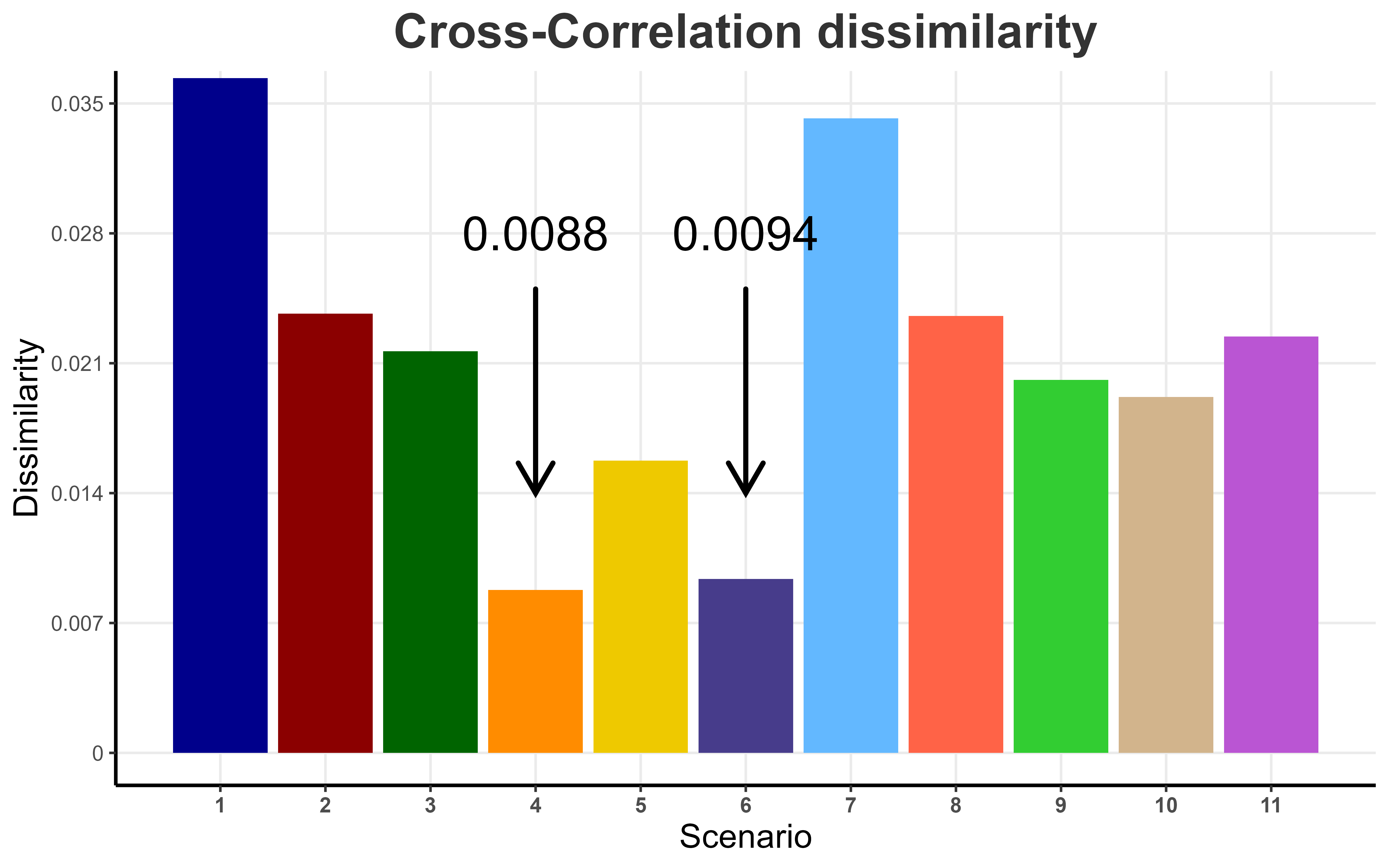}
\end{center}
\caption{$L_2$ (upper panel) and cross-correlation (lower panel) dissimilarities for each scenario using the method-1 statistic.}
\label{fig:meth1subsam}
\end{figure}

\begin{figure}[h!]
\begin{center}
\includegraphics[width=0.7\textwidth]{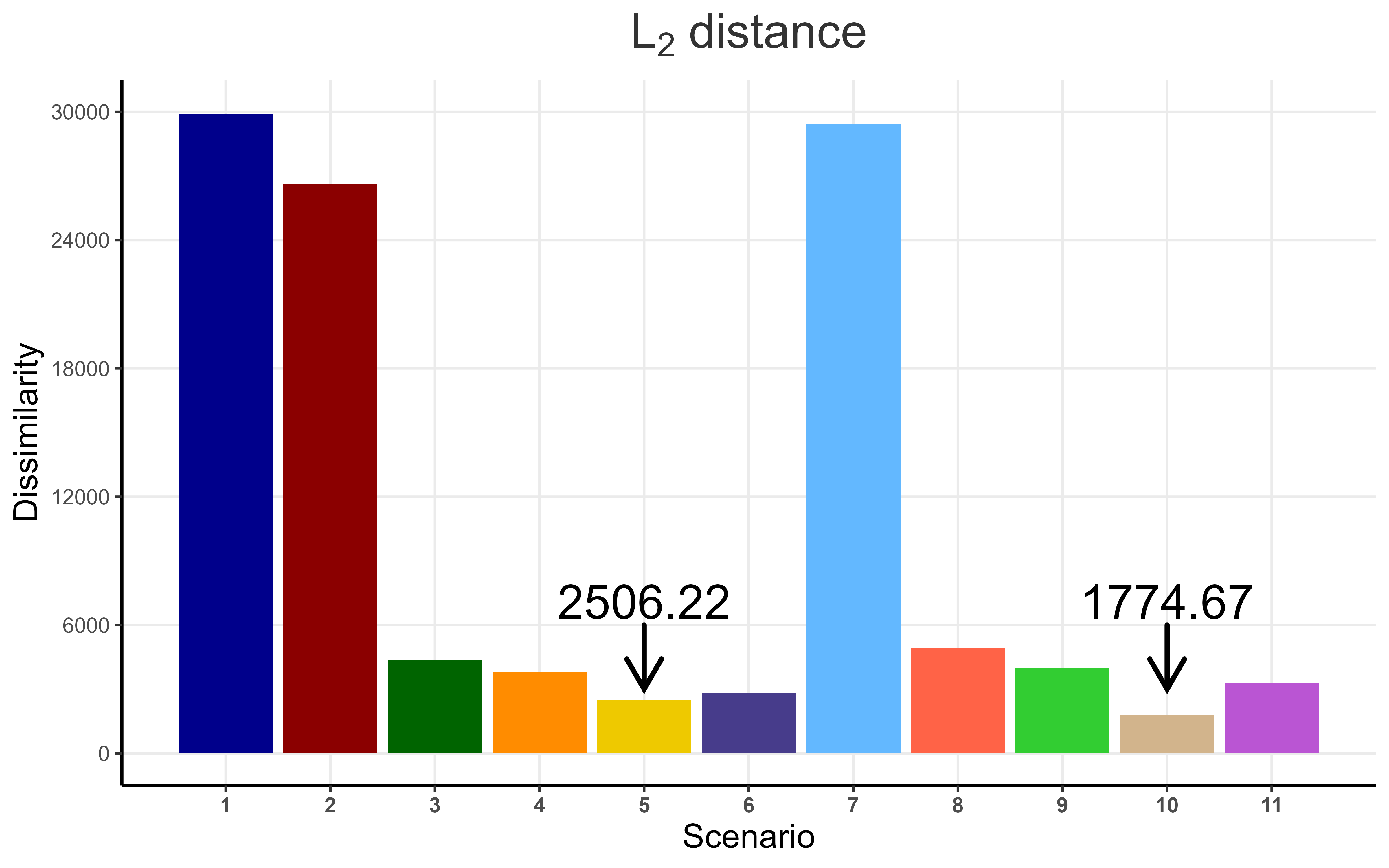}
\includegraphics[width=0.7\textwidth]{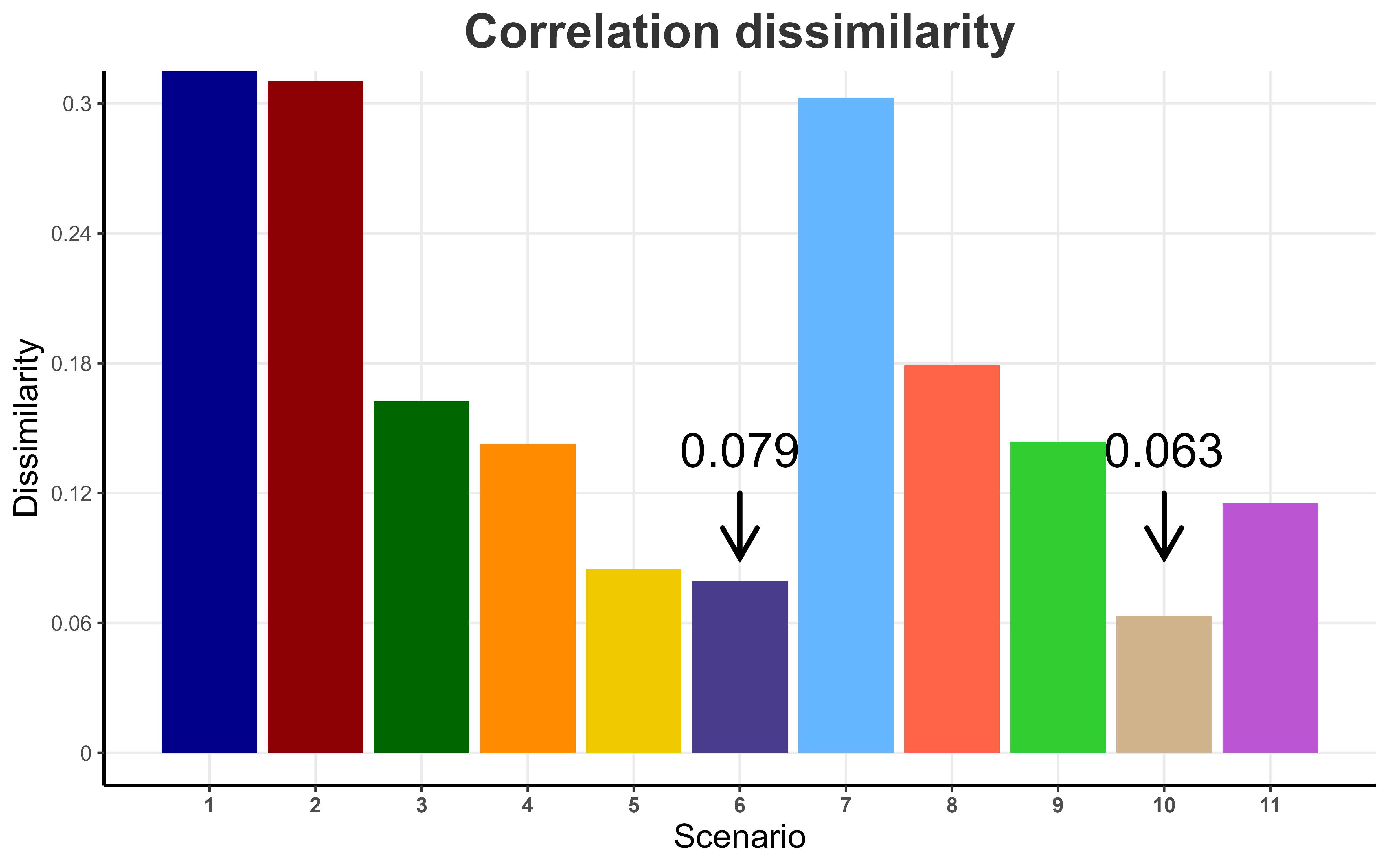}
\includegraphics[width=0.7\textwidth]{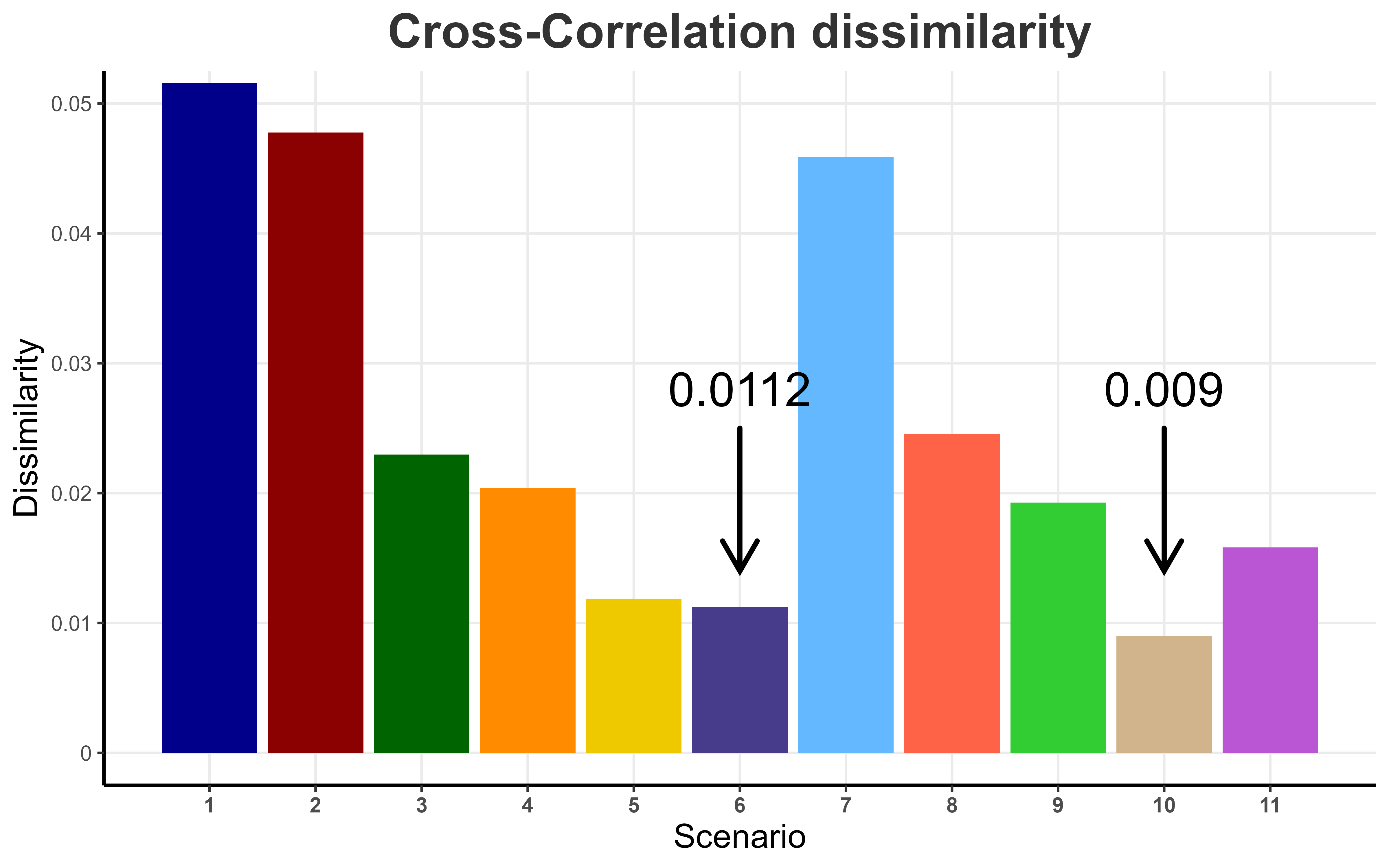}
\end{center}
\caption{$L_2$ (upper panel), correlation (middle panel) and cross-correlation (lower panel) dissimilarities for each scenario using the method-2 statistic.}
\label{fig:meth2subsam}
\end{figure}

\begin{figure}[h!]
\begin{center}
\includegraphics[width=0.9\textwidth]{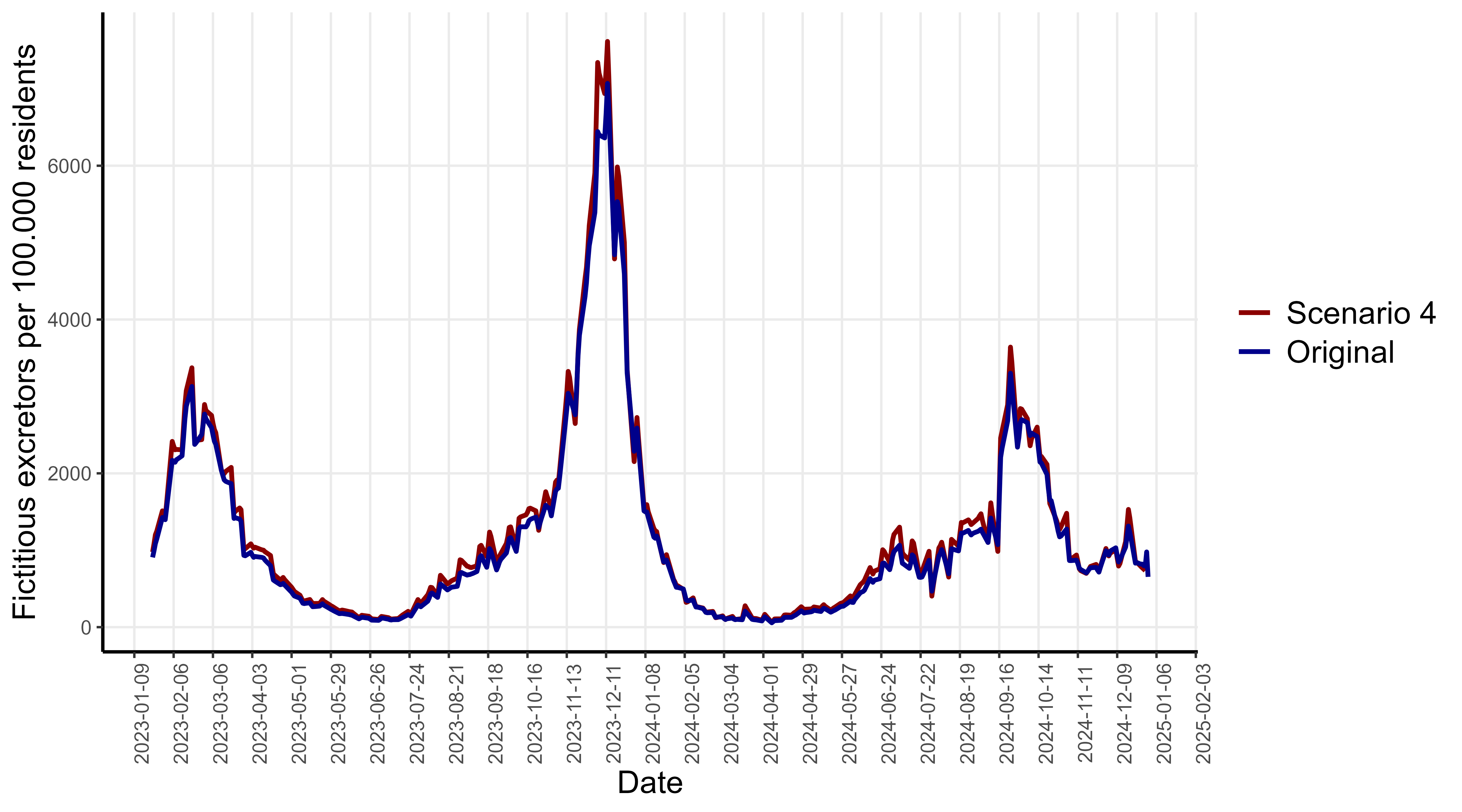}
\includegraphics[width=0.9\textwidth]{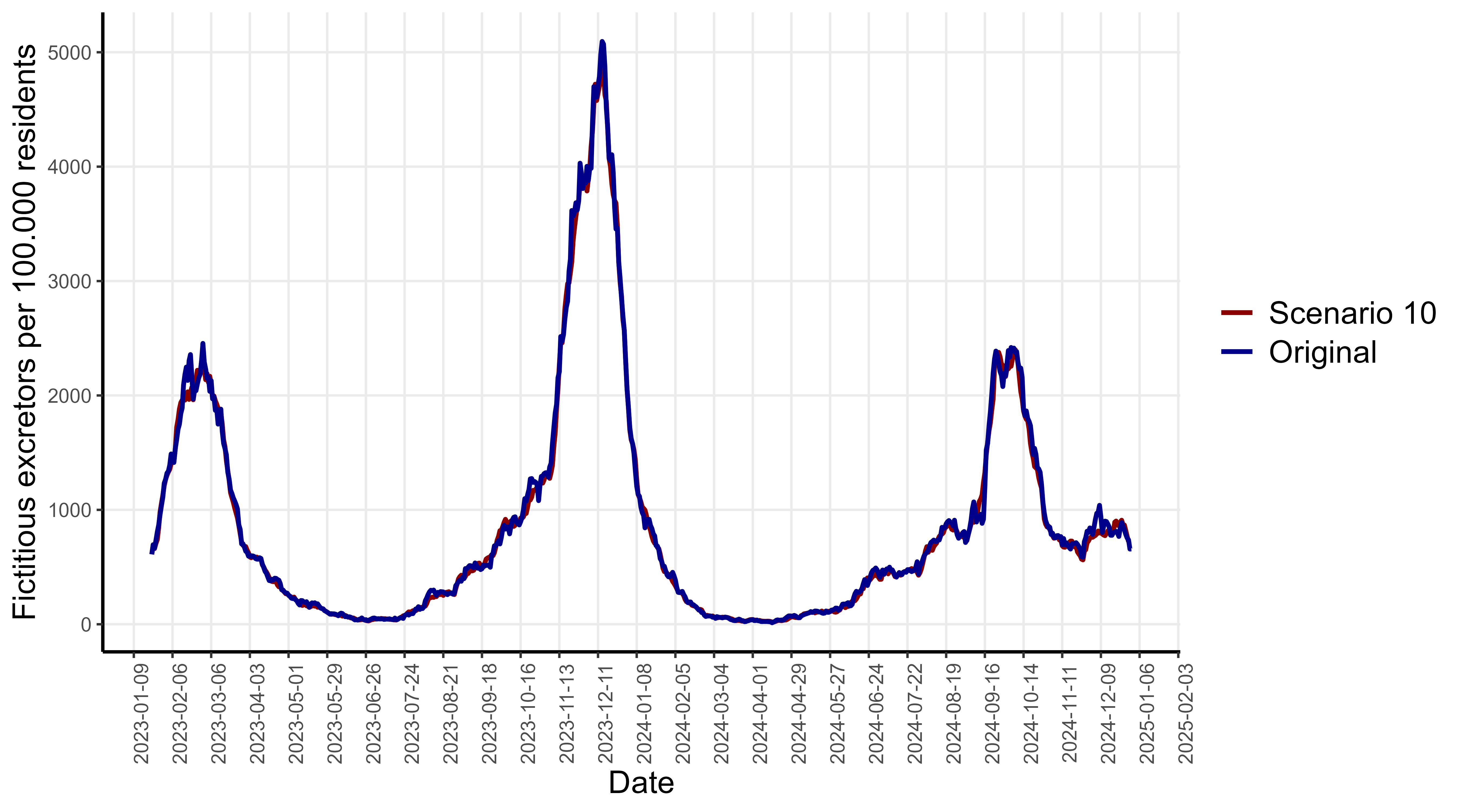}
\end{center}
\caption{The best scenario found for each method. Up: Method-1 statistic. Down: Method-2 statistic.}
\label{fig:bestscens}
\end{figure}

The scenarios we test by different combinations of sampling volume and frequency are shown in Table \ref{tab:subsampling}. In the main text, we show the results using the method-1 statistic and the correlation-based dissimilarity. In Figure \ref{fig:meth1subsam}, we show the $L_2$ and cross-correlation dissimilarities while, in Figure \ref{fig:meth2subsam}, we show the results for the method-2 statistic. In Figure \ref{fig:bestscens}, the best scenarios are depicted as time series.

\begin{table}[!h]
\caption{The scenarios tested for a different sampling program. The 6th scenario (S6) is found to be the best using the method-1 statistic and the 10th scenario (S10) is the best using the method-2 statistic.}
\label{tab:subsampling}
\begin{center}
\begin{tabular}{ c c c c c} 
\hline
& \multicolumn{4}{c}{Sampling volume} \\
Sampling frequency & 48 WWTPs & 24 WWTPs & 9 WWTPs & 1 WWTP \\ \hline
2 per week & Reference & S6 & S4 & S2 \\
1 per week & S10 & S5 & S3 & S1 \\
1 per 2 weeks & S11 & S9 & S8 & S7 \\
\hline
\end{tabular}
\end{center}
\end{table}

We apply the same method of estimating the dissimilarity between the original method-2 national curve and each of 8 scenarios of different sewer types and size. The scenarios can be seen in Table \ref{tab:sewertypesize}.

\begin{table}[!h]
\caption{The scenarios tested regarding sewer type and size.}
\label{tab:sewertypesize}
\begin{center}
\begin{tabular}{ c c c c c} 
\hline
& \multicolumn{4}{c}{Sewer type} \\
Sewer size & unknown & separate & combined & separate \& combined \\ \hline
$>$ 100,000 & S1 & S3 & S5 & S7 \\
$<$ 100,000 & S2 & S4 & S6 & S8 \\
\hline
\end{tabular}
\end{center}
\end{table}

\begin{figure}[h!]
\begin{center}
\includegraphics[width=0.9\textwidth]{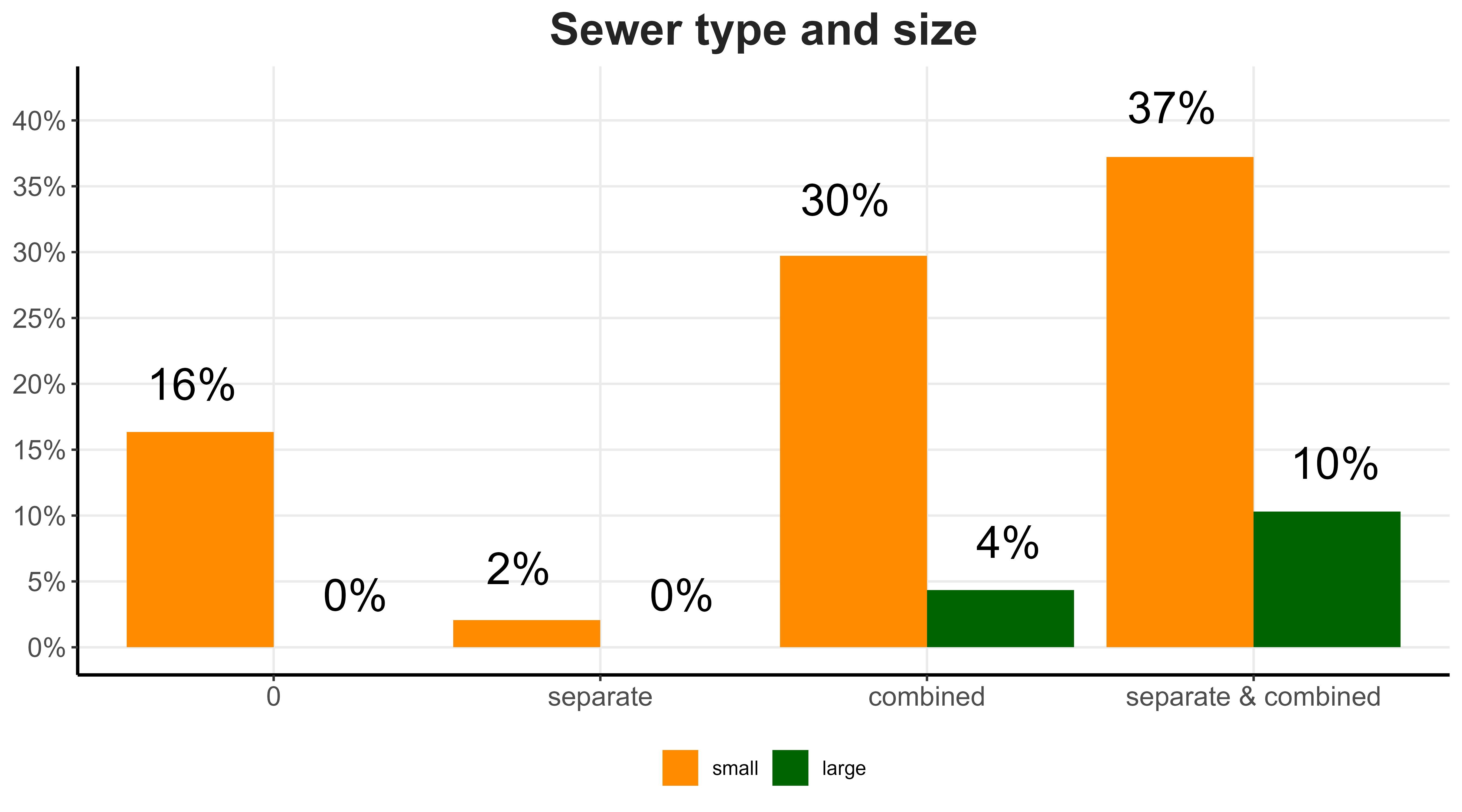}
\end{center}
\caption{Different combinations of sewer type and size and their proportions in the data.}
\label{fig:typesize}
\end{figure}

\begin{figure}[h!]
\begin{center}
\includegraphics[width=0.9\textwidth]{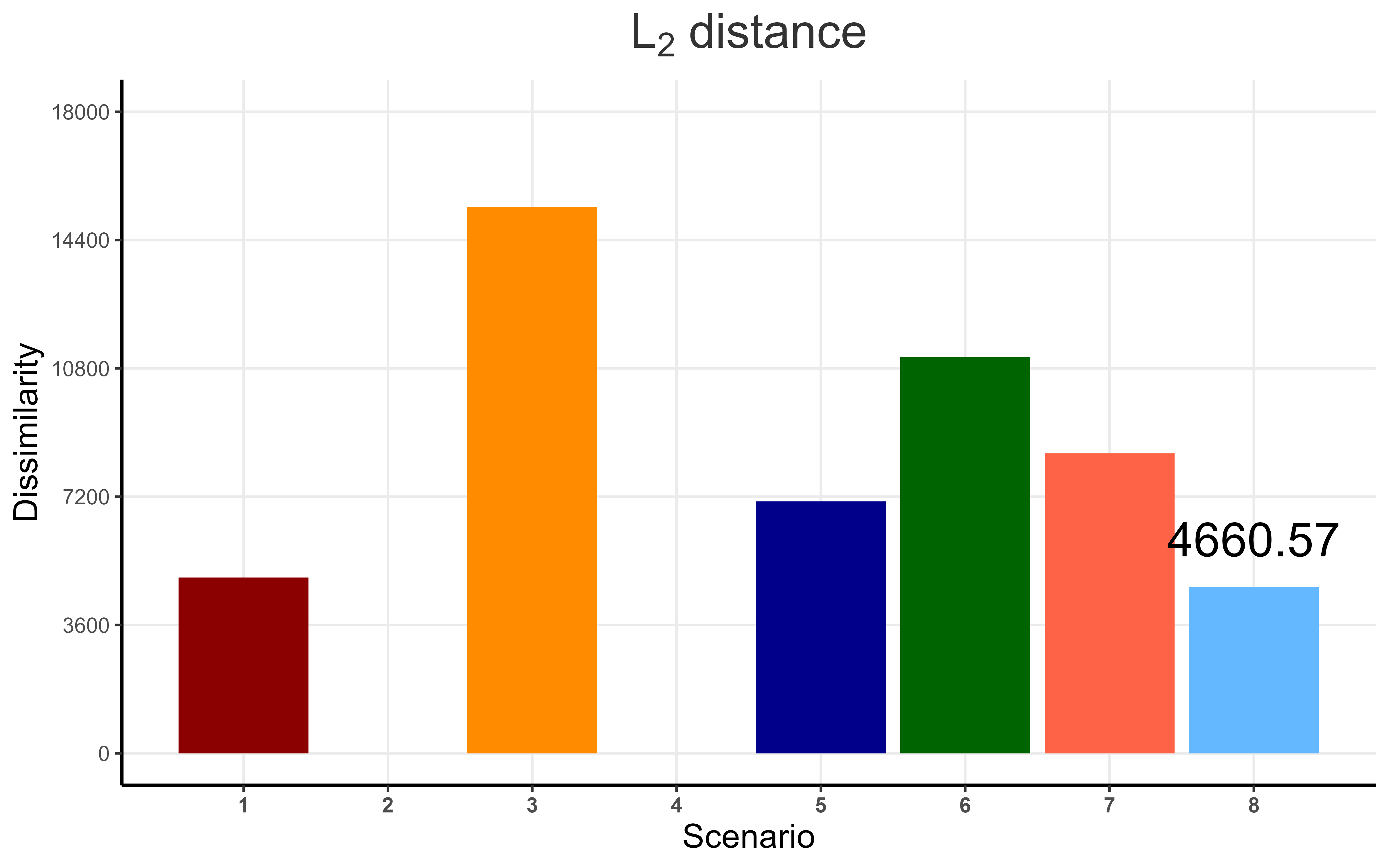}
\end{center}
\caption{The $L_2$ dissimilarity between each scenario of sewer type and size and the original method-2 statistic.}
\label{fig:l2dist_typesize}
\end{figure}

\begin{figure}[h!]
\begin{center}
\includegraphics[width=0.9\textwidth]{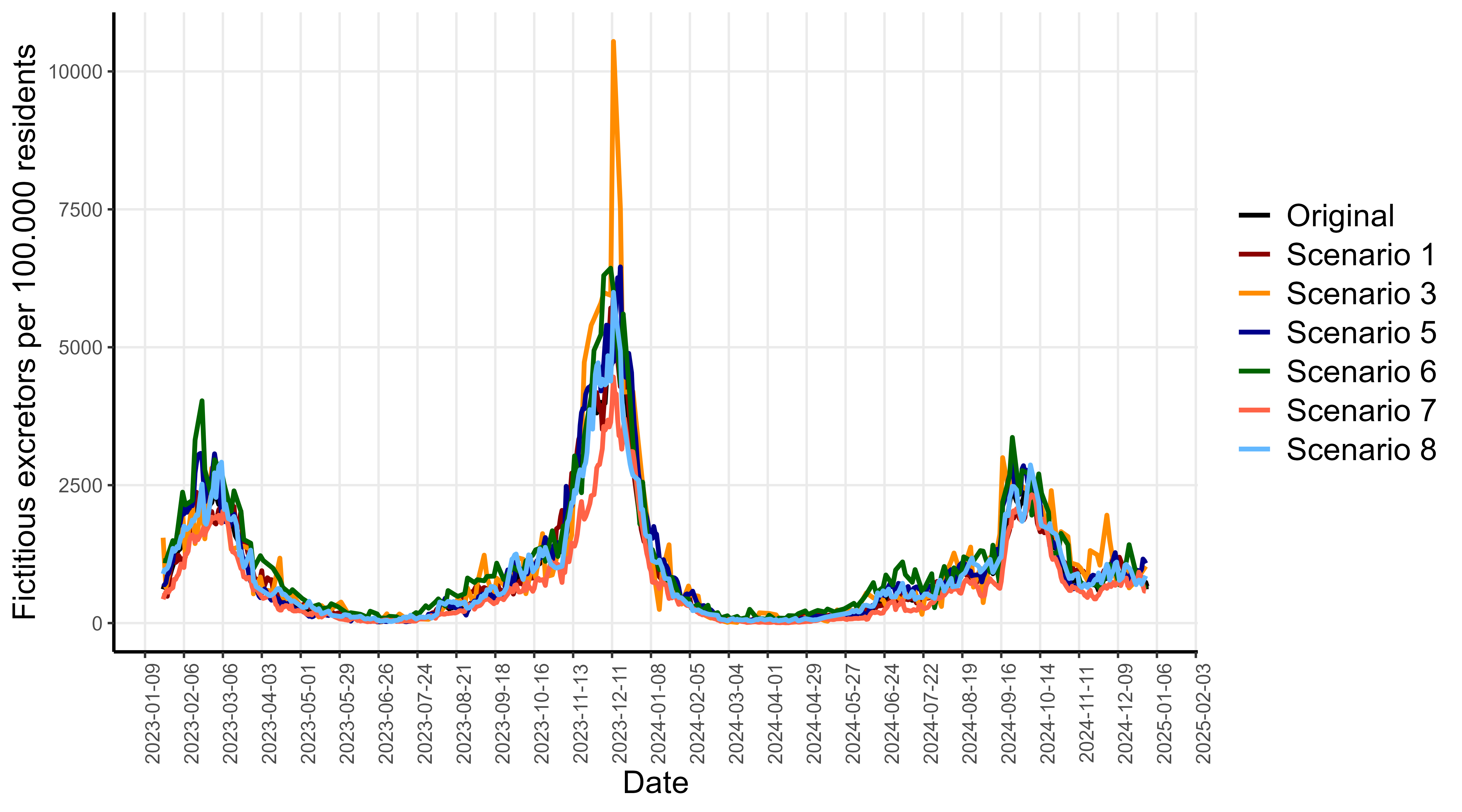}
\end{center}
\caption{The scenarios of sewer type and size as time series.}
\label{fig:scens_meth2}
\end{figure}

\section*{Appendix F. Statistical process monitoring results}

\begin{figure}[h!]
\begin{center}
\includegraphics[width=0.9\textwidth]{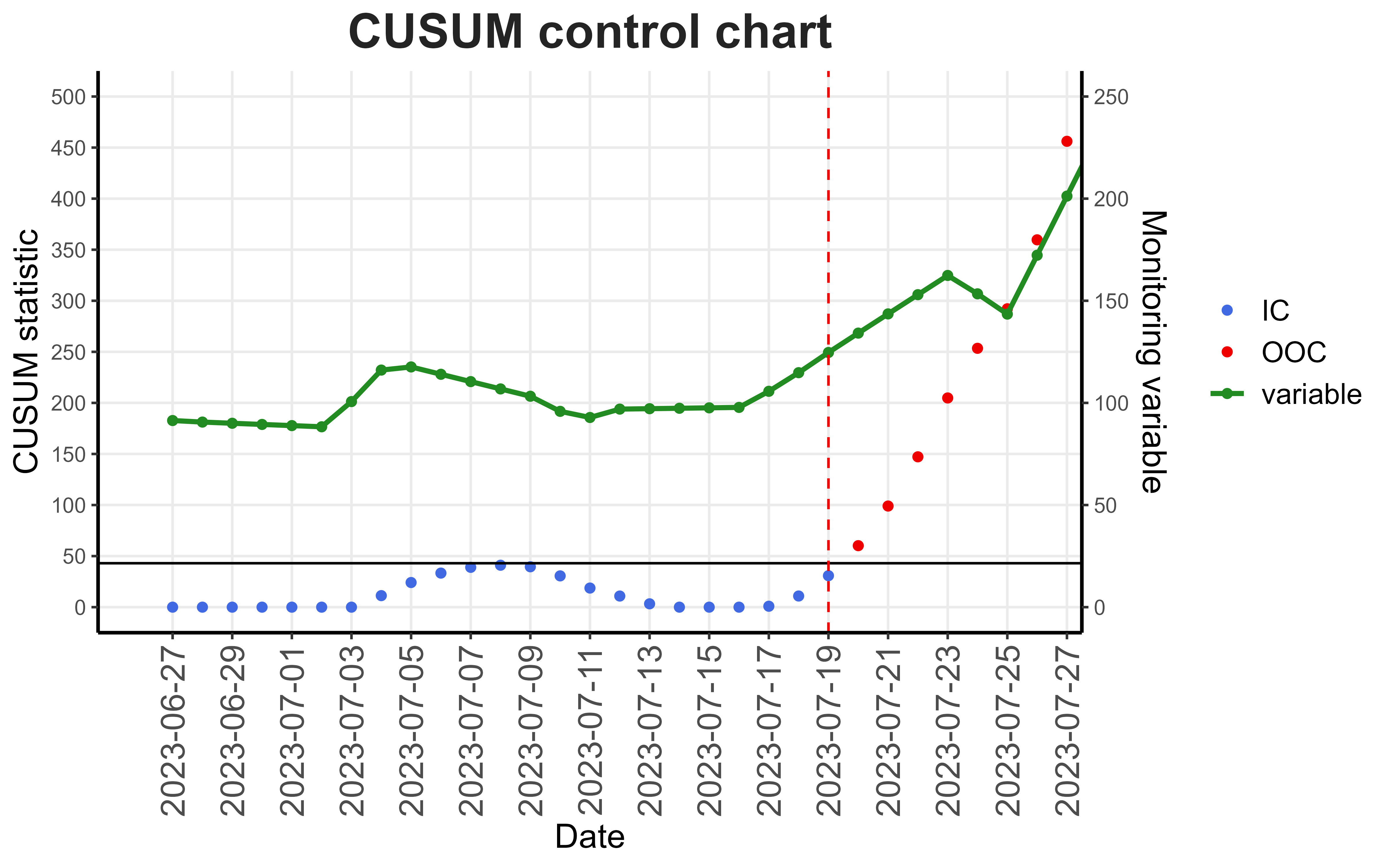}
\end{center}
\caption{The CUSUM chart detecting a shift from $\mu_0=100$ to $\mu_1=150$ gives $\mathrm{ARL}_1=1.18$. The IC period is from 27/06/2023 until 19/07/2023.}
\label{fig:orig_cusum_meth1_2}
\end{figure}

\begin{figure}[h!]
\begin{center}
\includegraphics[width=0.9\textwidth]{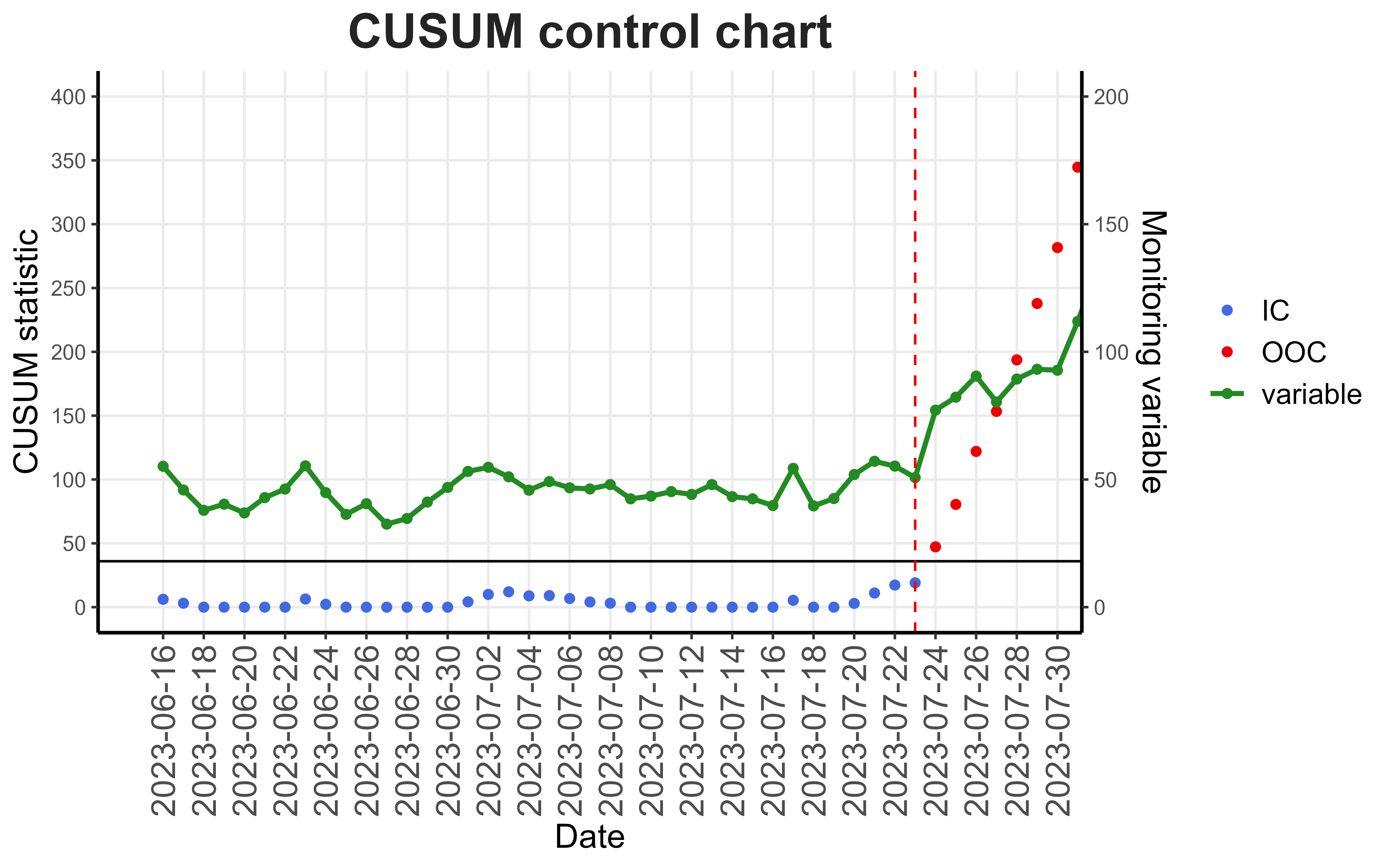}
\end{center}
\caption{The CUSUM chart detecting a shift from $\mu_0=45$ to $\mu_1=55$ gives $\mathrm{ARL}_0=564$ and $\mathrm{ARL}_1=4.76$. The IC period is from 16/06/2023 until 23/07/2023.}
\label{fig:orig_cusum_meth2_2}
\end{figure}

\begin{figure}[h!]
\begin{center}
\includegraphics[width=0.9\textwidth]{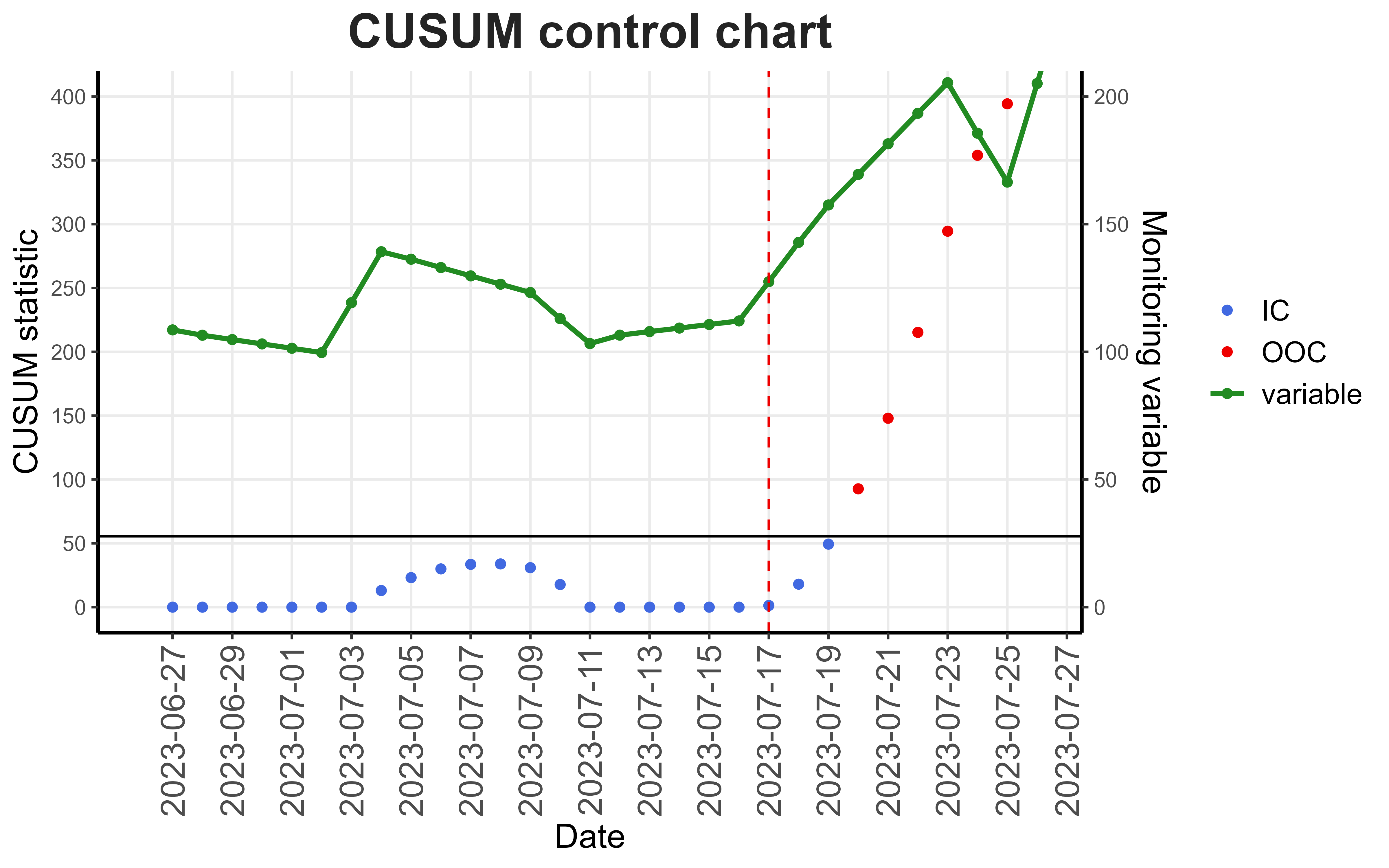}
\end{center}
\caption{The CUSUM chart detecting a shift from $\mu_0=120$ to $\mu_1=150$ gives $\mathrm{ARL}_0=564$ and $\mathrm{ARL}_1=2.8$. The IC period is from 27/06/2023 17/07/2023.}
\label{fig:redu_cusum_2}
\end{figure}

\begin{figure}[h!]
\begin{center}
\includegraphics[width=0.9\textwidth]{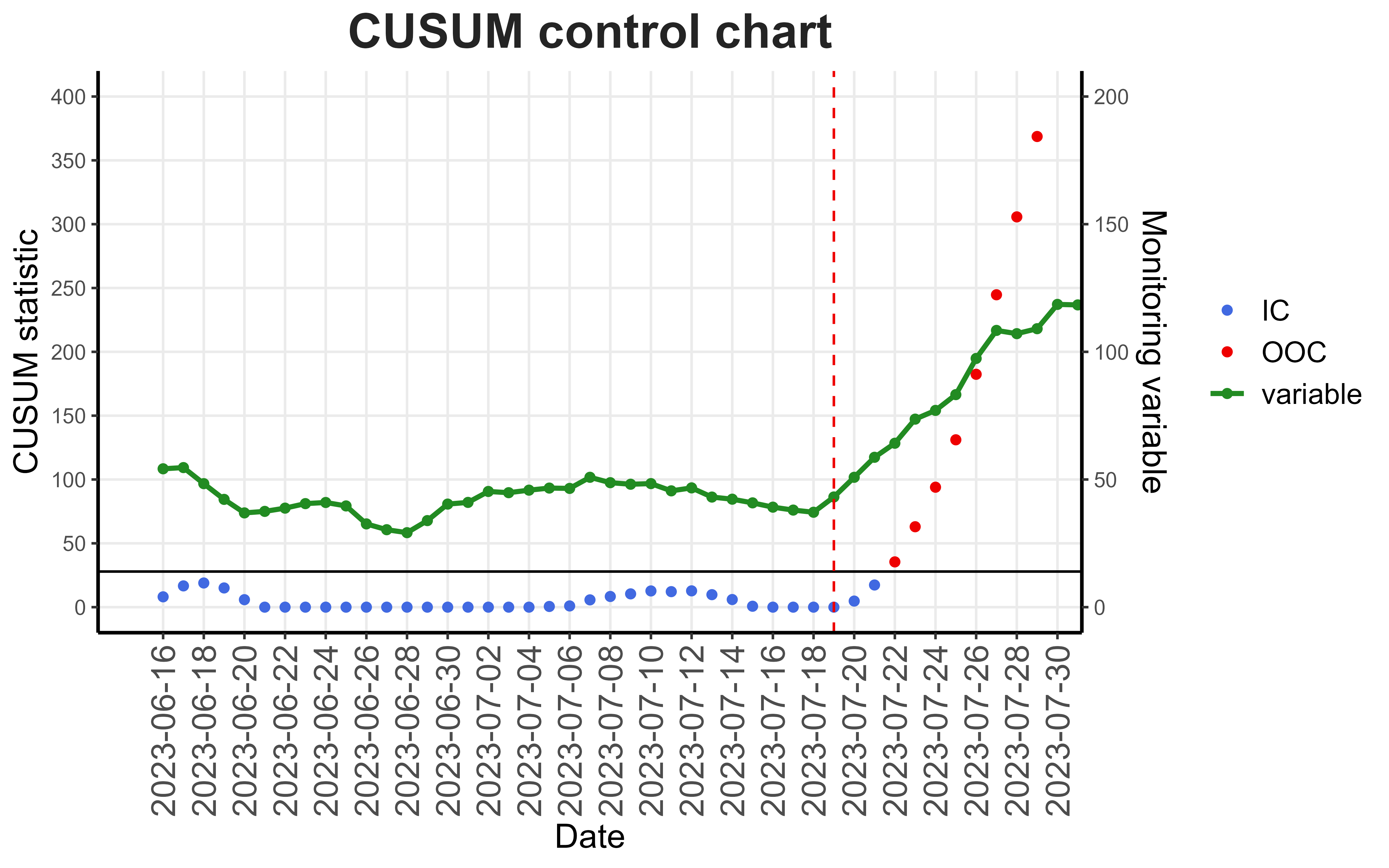}
\end{center}
\caption{The CUSUM chart detecting a shift from $\mu_0=43$ to $\mu_1=47$ gives $\mathrm{ARL}_0=564$ and $\mathrm{ARL}_1=19.89$. The IC period is from 16/06/2023 until 19/07/2023.}
\label{fig:redu_cusum_meth2_2}
\end{figure}

The one-sided CUSUM chart uses the statistic $C_i^+=\max{0,x_i-(\mu_0+K)+C_{i-1}^+}$, where $K=k\sigma$ and the decision interval is $H=h\sigma$ with $\sigma$ being the standard deviation of the process. We set the parameters $k=1/2$ and $h=4.5$ so that all the following charts have similar $\mathrm{ARL}$ performance. Here, $\mathrm{ARL}_0=564$. The proof-of-concept nature of the approach is indicated by the fact that the in-control (IC) and out-of-control (OOC) periods and mean values $mu_0$ and $\mu_1$ cannot be objectively defined, but this is an issue that can be fixed by an appropriate experiment and more data. We test the cases where the monitoring statistic is either the method-1 statistic, or the `reduced' method-1 statistic, or the method-2 statistic, or the `reduced' method-2 statistic. `Reduced' means that we use the data from the best sub-sampling scenario (scenario 6 for method-1 and scenario 10 for method-2). The $\mathrm{ARL}$ is calculated using the Sigmund's formula. The results are depicted in Figures \ref{fig:orig_cusum_meth1_2}, \ref{fig:orig_cusum_meth2_2}, \ref{fig:redu_cusum_2}, \ref{fig:redu_cusum_meth2_2}. The green points indicate the monitoring variable of the fictitious excretors, the red vertical line indicate the end of the IC period. The blue and red points are in- and out-of-control points of the CUSUM statistic.

\begin{figure}[h!]
\begin{center}
\includegraphics[width=0.9\textwidth]{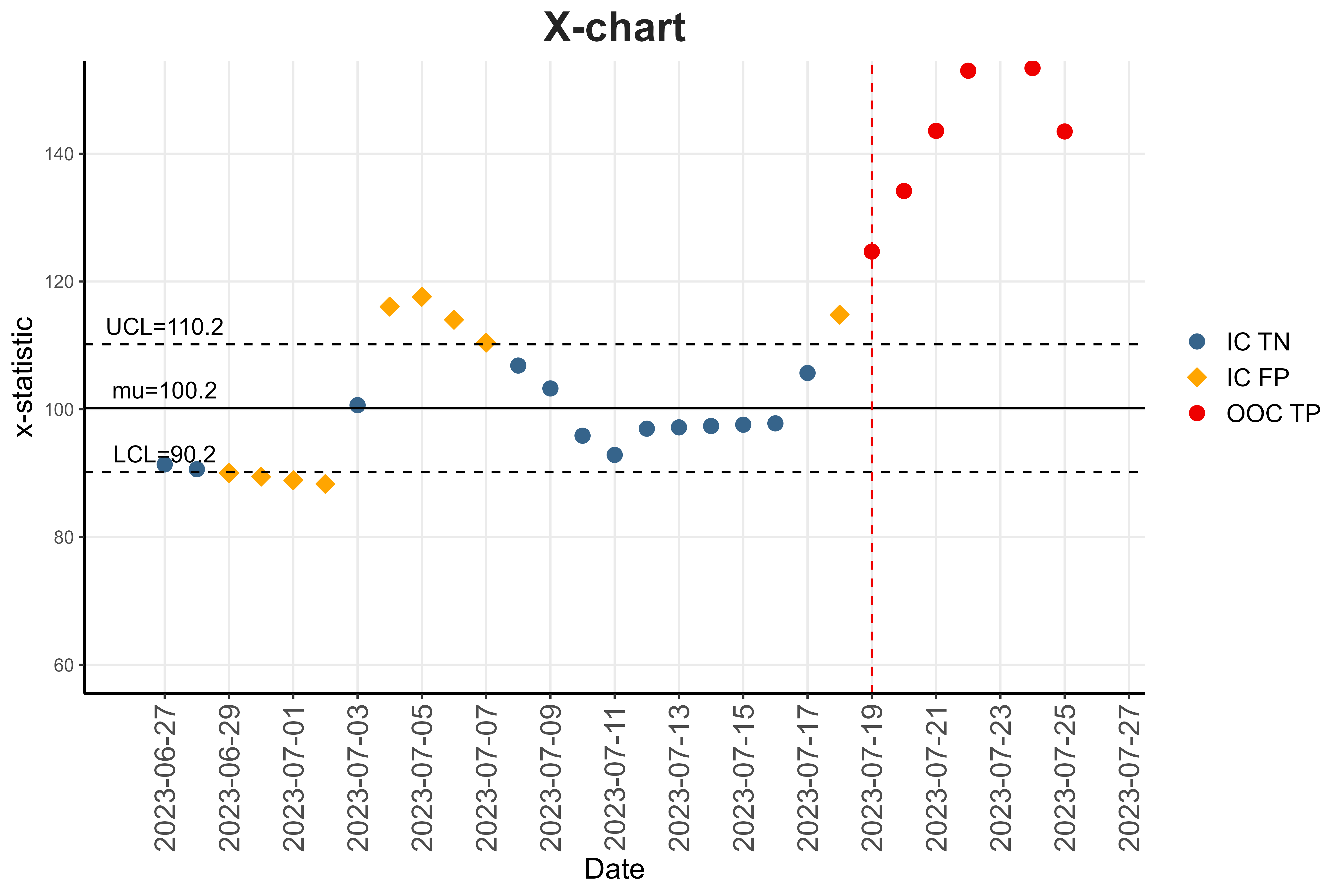}
\end{center}
\caption{The $X$ chart for the method-1 statistic. TN stands for true negative, FP stands for false positice and TP stands for true positive.}
\label{fig:orig_shew_noaut}
\end{figure}

\begin{figure}[h!]
\begin{center}
\includegraphics[width=0.9\textwidth]{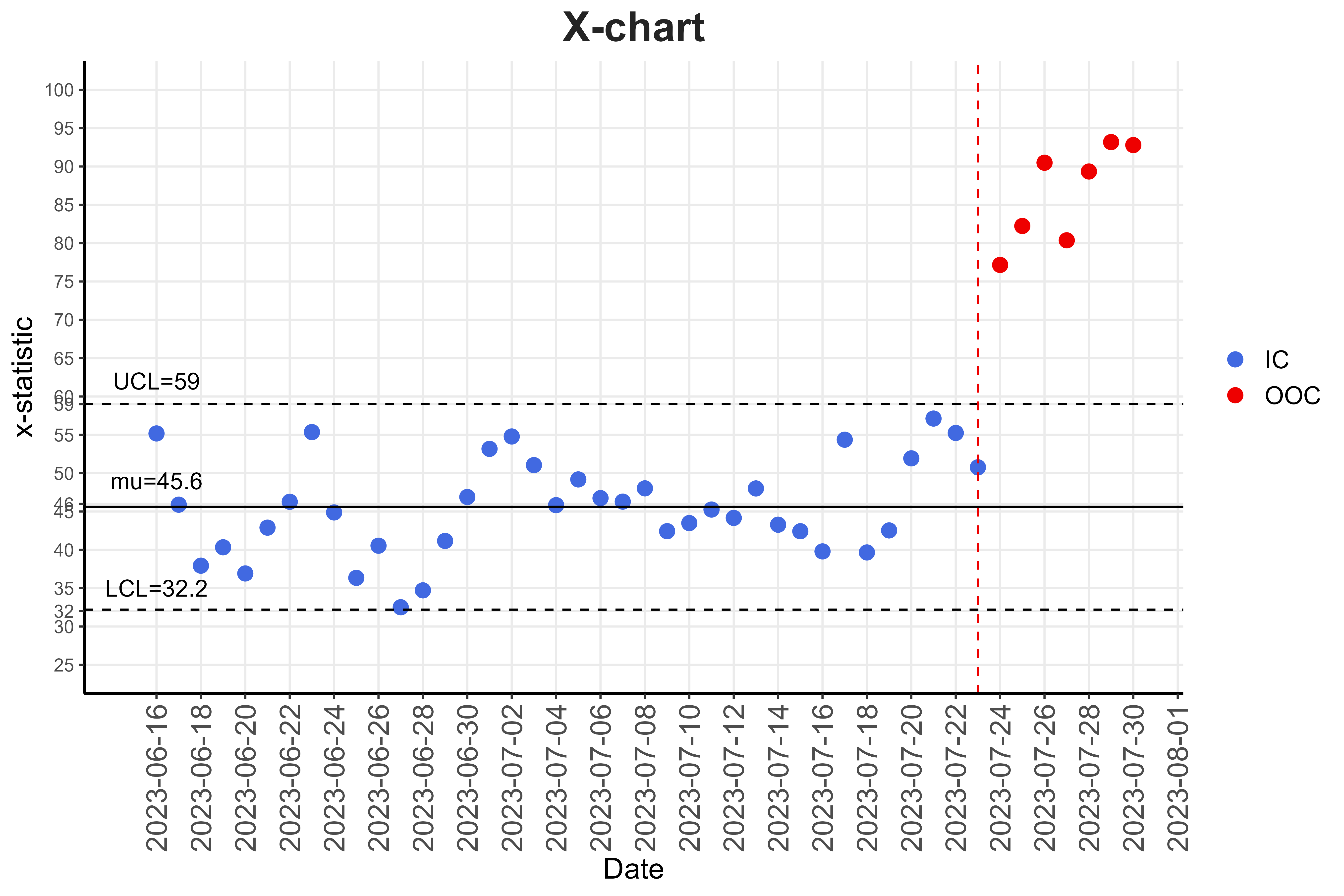}
\end{center}
\caption{The $X$-chart for the method-2 statistic.}
\label{fig:orig_shew_noaut2}
\end{figure}

\begin{figure}[h!]
\begin{center}
\includegraphics[width=0.9\textwidth]{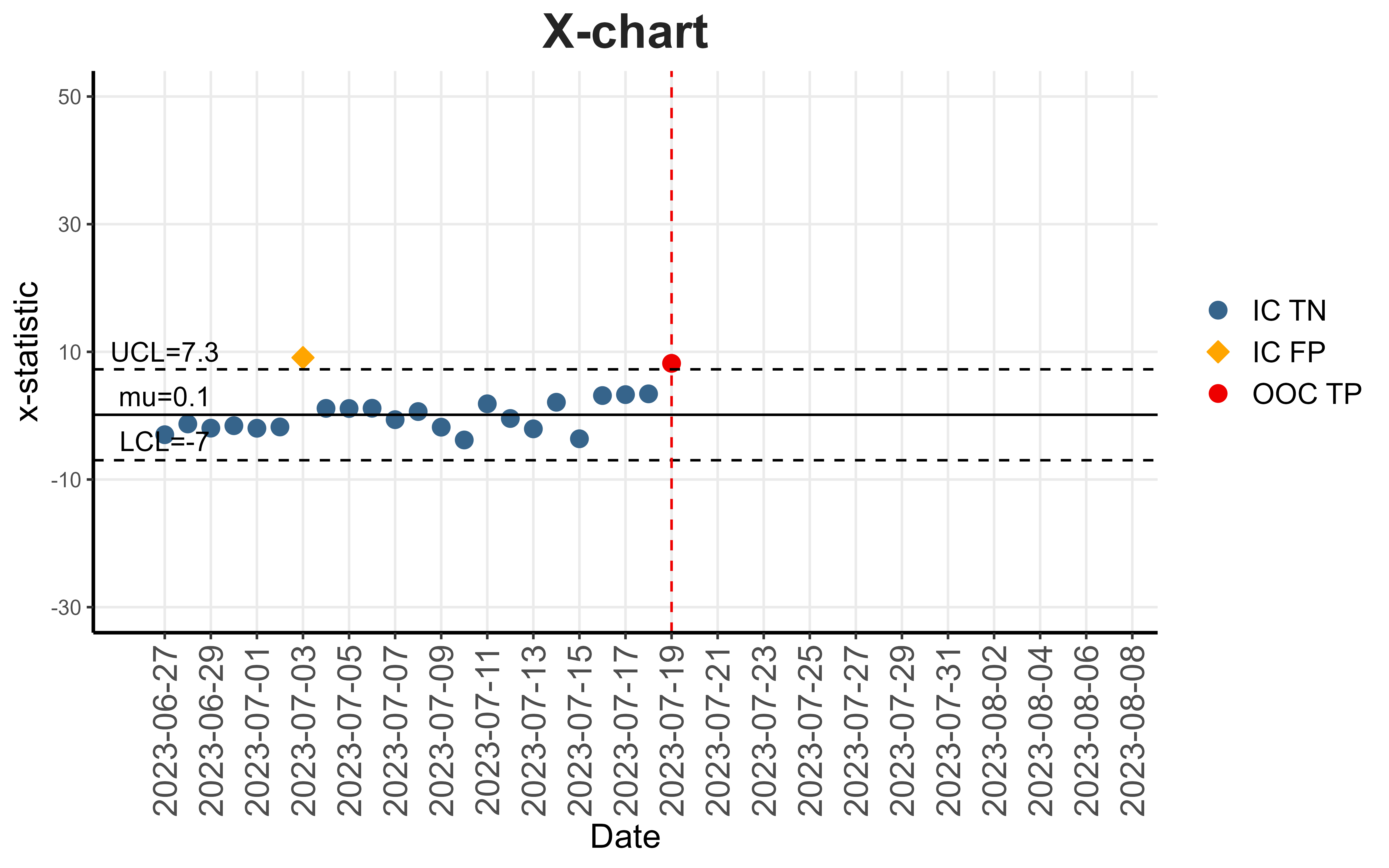}
\end{center}
\caption{The $X$ chart for the method-1 statistic. `TN' stands for `true negative', `FP' stands for `false positive' and `TP' stands for `true positive'.}
\label{fig:orig_shew_aut2}
\end{figure}

Another type of charts tested is the Shewhart $X$-chart, seen in Figures \ref{fig:orig_shew_noaut} and \ref{fig:orig_shew_noaut2} for the original data of the method-1 and method-2 statistics respectively. The method does not seem to work well for method-1, because of the inherent autocorrelation (Box-Ljung test p-value 0.0001). Thus, we try to fix this issue by training an $\mathrm{ARIMA}$ model and then fit the chart on the residuals. We fit an $\mathrm{ARIMA}(1,0,3)$ and the results are shown in Figure \ref{fig:orig_shew_aut2}, where the performance has been clearly improved.

\bibliographystyle{apa}
\bibliography{bibtex}